\documentclass[useAMS,usenatbib]{mn2e}

\usepackage{psfig}
\usepackage{amsmath}
\usepackage{amssymb}
\usepackage{upgreek}
\usepackage{longtable}
\usepackage[mathscr]{eucal}
\usepackage[dvips]{graphicx}
\usepackage{float}

\title
[Kinematics of the Sco-Cen OB association]
{The Kinematics of the Scorpius-Centaurus OB Association from {\it Gaia} DR1}
\author
[Wright et al.]
{Nicholas J. Wright$^{1}$ and Eric E. Mamajek$^{2,3}$\\
$^{1}$Astrophysics Group, Keele University, Keele, ST5 5BG, UK\\
$^{2}$Jet Propulsion Laboratory, California Institute of Technology, Pasadena, CA 91109, USA\\
$^{3}$Department of Physics \& Astronomy, University of Rochester, 500 Wilson Blvd., Rochester, NY 14627-0171, USA
}

\begin{document}
\maketitle

\begin{abstract}

We present a kinematic study of the Scorpius-Centaurus (Sco-Cen) OB association (Sco OB2) using {\it Gaia} DR1 parallaxes and proper motions. Our goal is to test the classical theory that OB associations are the expanded remnants of dense and compact star clusters disrupted by processes such as residual gas expulsion. {\it Gaia} astrometry is available for 258 out of 433 members of the association, with revised {\it Hipparcos} astrometry used for the remainder. We use this data to confirm that the three subgroups of Sco-Cen are gravitationally unbound and have non-isotropic velocity dispersions, suggesting they have not had time to dynamically relax. We also explore the internal kinematics of the subgroups to search for evidence of expansion. We test Blaauw's classical linear model of expansion, search for velocity trends along the Galactic axes, compare the expanding and non-expanding convergence points, perform traceback analysis assuming both linear trajectories and using an epicycle approximation, and assess the evidence for expansion in proper motions corrected for virtual expansion / contraction. None of these methods provide coherent evidence for expansion of the subgroups, with no evidence to suggest that the subgroups had a more compact configuration in the past. We find evidence for kinematic substructure within the subgroups that supports the view that they were not formed by the disruption of individual star clusters. We conclude that Sco-Cen was likely born highly substructured, with multiple small-scale star formation events contributing to the overall OB association, and not as single, monolithic bursts of clustered star formation.

\end{abstract}

\begin{keywords}

stars: formation - stars: kinematics and dynamics - open clusters and associations: individual: Scorpius-Centaurus, Sco OB2, Upper Scorpius, Upper Centaurus-Lupus, Lower Centaurus-Crux.

\end{keywords}

\section{Introduction}

\begin{figure*}
\begin{center}
\includegraphics[width=510pt]{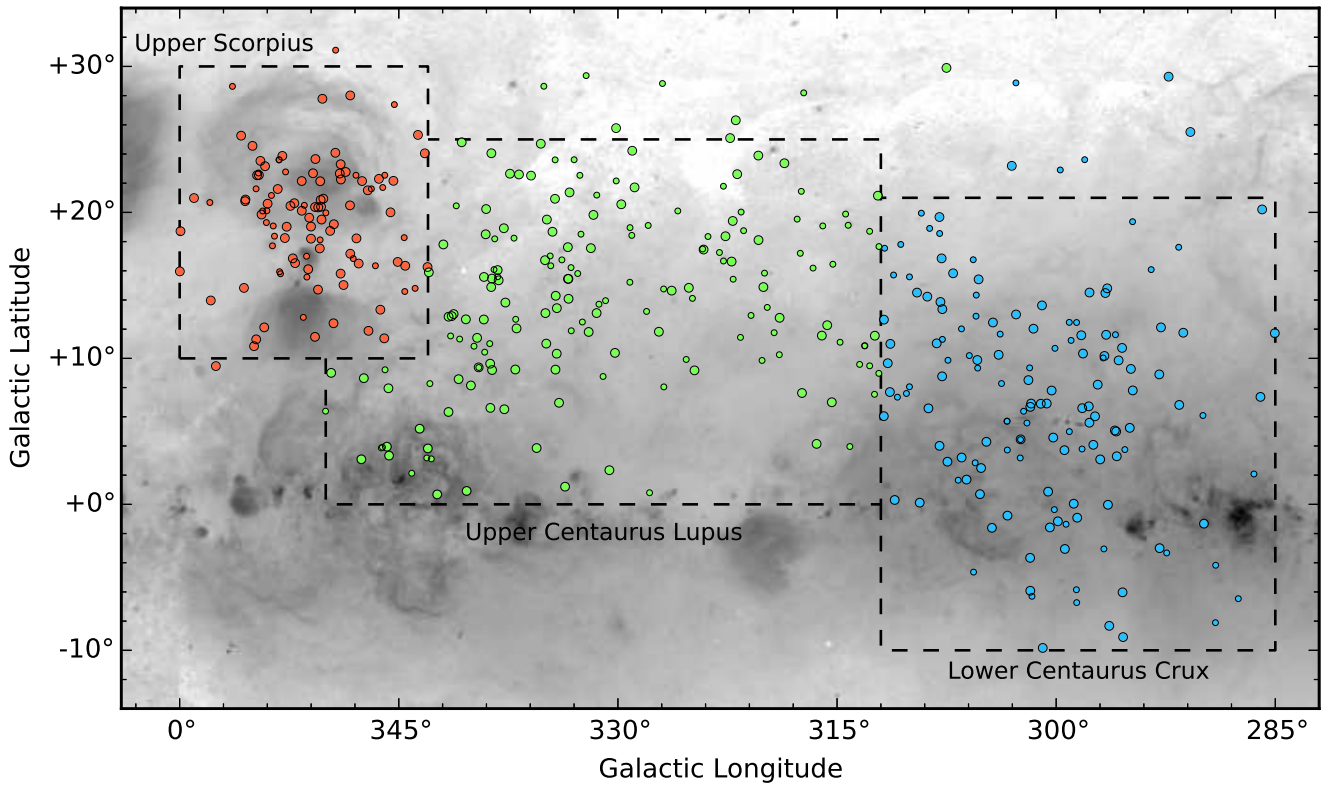}
\caption{Spatial distribution of Sco-Cen members \citep{rizz11} projected onto an inverted H$\alpha$ image from \citet{fink03}. Large circles represent sources with {\it Gaia} DR1 proper motions and parallaxes, while small circles are sources not detected by {\it Gaia} for which {\it Hipparcos} proper motions and parallaxes were used. The dashed lines show the division of the association into the three subgroups according to \citet{deze99}, with the symbols coloured according to which of the three subgroups they nominally belong: red for US, green for UCL, and blue for LCC. Sources outside of these boundaries have been assigned to the nearest subgroup.}
\label{spatial_distribution}
\end{center}
\end{figure*}

The environment where stars form and spend the first few million years of their lives has profound consequences for the rest of their lives and their potential to host habitable exoplanets. Ultraviolet (UV) radiation from nearby massive stars can lead to photoevaporation of protoplanetary disks \citep[e.g.,][]{odel94,wrig12a,guar16}, while close encounters in dense clusters can affect the properties of binary and multiple systems \citep{krou99,mark12,park12d}, protoplanetary disks \citep{scal01,olcz08,roso14}, and influence the formation of planetary systems \citep{adam06,park12c}. Stars born in low-density environments \citep[e.g.,][]{bres10,wrig14b} where dynamical interactions are rare and UV radiation fields are weaker may form planetary and binary systems with little or no external disruption.

The majority of young stars are observed in groups or clusters of some sort, but by an age of 10~Myr only $\sim$10\% of stars are found in bound clusters \citep{lada03}. The classical explanation for this is that star clusters form embedded within molecular clouds, held together by the gravitational potential of both the stars and the gas, but when feedback disperses the gas left over from star formation \citep[which can account for well over half the mass of the system,][]{elme00} then the cluster becomes supervirial and will expand and disperse \citep[e.g.,][]{tutu78,hill80,lada84,baum07}. During the expansion the system would briefly be visible as a low-density group of young stars known as an {\it association} \citep[e.g.,][]{amba47,blaa64,brow97,krou01}, before dispersing into the Galactic field.

This theoretical framework has existed for many decades, and while many observations support aspects of this model \citep[e.g.,][and references therein]{lada03} it has been difficult to test kinematically due to the lack of high-precision proper motions necessary to assess evidence for expansion. \citet{wrig16} performed the first such study, using ground-based proper motions to study the kinematics of the Cygnus~OB2 association. They found that the association lacked the coherent radial expansion pattern expected if it was an expanding star cluster. Its non-isotropic velocity dispersions and both physical \citep{wrig14b} and kinematic substructure instead argued for an origin as a highly substructured and extended association of stars.

The recent release of data from the {\it Gaia} satellite \citep{prus16} is set to change this picture, providing orders of magnitude improvement in the quality and quantity of proper motions. In this study we use {\it Gaia} astrometry to study the kinematics of the high- and intermediate-mass stars in the Sco-Cen OB association to assess its dynamical state, search for evidence of expansion, and constrain its initial conditions.

Sco-Cen (also known as Sco OB2) is the nearest OB association to the Sun and therefore the nearest region that has recently formed massive stars. Spanning distances of $\sim$100--150~pc \citep{deze99} and being at least 100~pc across, the association spans almost 90 degrees on the sky (see Figure~\ref{spatial_distribution}). The association consists of three subgroups \citep[first defined by][]{blaa46} known as Upper Scorpius (US), Upper Centaurus-Lupus (UCL) and Lower Centaurus-Crux (LCC), with median ages of 11, 16, and 17 Myr respectively, though each has a considerable spread of ages \citep{peca16}. Because of its age the region has likely already seen several supernovae \citep[e.g.,][]{brei16}, one of which may have been responsible for the ejected runaway star $\zeta$~Oph \citep{dege92,hoog00}. 

The association has been the focus of numerous studies of its low-mass population \citep[e.g.,][]{prei99,mama02}, with studies focussing on the binary properties \citep[e.g.,][]{kouw07,jans13}, circumstellar disks \citep[e.g.,][]{carp06,chen11}, stellar ages \citep[e.g.,][]{peca12,peca16}, eclipsing binaries \citep[e.g.,][]{krau15}, and low-mass stars and brown dwarfs \citep[e.g.,][]{lodi13}. Its high-mass population has been studied by numerous authors, including comprehensive studies by \citet{blaa46,blaa64b} and later, using {\it Hipparcos} astrometry, by \citet{deze99}, \citet{debr99}, and \citet{rizz11}.

This paper is outlined as follows. In Section 2 we present the data used. In Section 3 we recalculate the distance to the subgroups and quantify their 3D structure. In Section 4 we use proper motions and radial velocities (RVs) to measure the 3D velocity dispersions and bulk motions of the subgroups. In Section 5 we assess evidence for the expansion of the association using various methods, and in Section 6 we study its kinematic substructure. Finally in Section 7 we discuss our results and their implications.

\section{Observational data}

The sample of Sco-Cen members used for our kinematic study is based on the Bayesian membership analysis of \citet{rizz11}, who used the updated {\it Hipparcos} catalogue \citep{vanl07} to revise the kinematic selection of \citet{deze99}. Their linear model exploits positions, proper motions and parallaxes, as well as RVs from the 2nd version of the Catalogue of Radial Velocities with Astrometric Data \citep[CRVAD-2,][]{khar07}\footnote{\citet{murp15} note that some of the RVs used by \citet{rizz11} are non-spectroscopic `astrometric' RVs that are inappropriate for calculating space motions or membership. We investigated the phase-space distribution of these sources and found that their proper motions and parallaxes are still consistent with being members of Sco-Cen and so have retained them.} to produce a list of 436 high- and intermediate-mass members of Sco-Cen (the majority of which are of A \& B spectral types). Many of the sources discarded by \citet{rizz11} were rejected based on their parallax or RVs, two parameters that were not used by \citet{deze99} in their original membership selection. 

We refined this sample based on recent studies that suggest some of these stars are unlikely to be members of Sco-Cen. \citet{chen11} identify HIP~62428 as an F0III star whose position in the Hertzsprung-Russell diagram is inconsistent with being a member of Sco-Cen, while HIP~70833 and HIP~75824 exhibit low lithium equivalent widths that imply they are older than typical Sco-Cen stars (in the case of HIP~70833 the lithium is measured in its K-type binary companion). The vast majority of stars identified as non-members by \citet{mama02}, \citet{chen11} and \citet{peca12} that were included in the original sample of \citet{deze99} were not selected by \citet{rizz11} as members, reinforcing the validity of their membership selection. \citet{peca12} recommend excluding a number of stars based on their kinematics, one of which is retained in the \citet{rizz11} catalogue, HIP~56227. We choose to retain this star for the time being and minimise the impact kinematic selection may have on our sample.

This leaves us with a membership list of 433 stars (after removing 3 stars), 107 in US, 179 in UCL, and 147 in LCC, as shown in Figure~\ref{spatial_distribution}. For the handful of stars that cannot be assigned to a subgroup based on the division of \citet{deze99} we put them in the nearest subgroup on the plane of the sky. While \citet{rizz11} concluded that the strict boundaries between the three subgroups are somewhat arbitrary we continue to use them here for the ease of analysing such an extended OB association and to allow comparison with previous studies. Due to the current lack of high-precision astrometry for most lower-mass members of the association and to avoid being spatially biased by the various pointings used for the spectroscopic observations of the low-mass stars we limit our sample in this study to the high-mass members observed by {\it Hipparcos} and {\it Gaia}.

\subsection{{\it Gaia} and {\it Hipparcos} astrometry}

\begin{figure*}
\begin{center}
\includegraphics[height=510pt, angle=270]{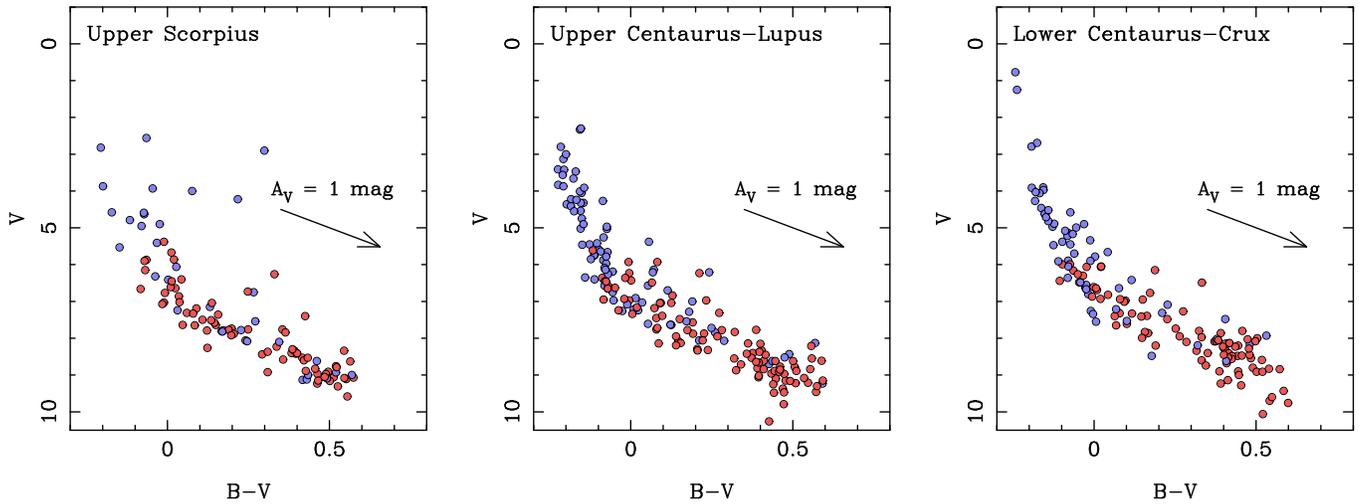}
\caption{Colour-magnitude diagrams for our Sco-Cen samples illustrating the completeness of the {\it Gaia} data \citep[photometry from][]{ande12}. Sources with astrometry from {\it Gaia} are shown in red and those for which {\it Hipparcos} data was used are shown in blue. A 1 magnitude reddening vector is show for illustrative purposes. The mean extinctions of members in each subgroup are $A_V$ = 0.73, 0.17, and 0.23 mag for US, UCL and LCC, respectively \citep{dege89}.}
\label{cmd}
\end{center}
\end{figure*}

Astrometric data for our sample comes from {\it Gaia} \citep{prus16} data release 1 \citep[DR1,][]{brow16}, which contains results from the Tycho-{\it Gaia} Astrometric Solution \citep[TGAS,][]{mich15,lind16} that combines astrometry from {\it Gaia}, {\it Hipparcos} \citep{esa97} and Tycho-2 \citep{hog00}. Of particular relevance for our sample is the subset of 93,635 stars in TGAS where {\it Hipparcos} positions at epoch J1991.25 were combined with {\it Gaia} observations from the first 14 months of the mission (2014 to 2015). The parallaxes and proper motions calculated from this data are of particularly high precision, with uncertainties of 0.3~mas and 0.06~mas/yr, respectively, though additional systematic uncertainties are still present \citep{brow16,lind16}. This data has some limitations since the astrometric solutions were calculated assuming single star behaviour (i.e., no consideration was made for the sources being binary systems) and perspective acceleration was not accounted for \citep{lind16}, but despite this the data should be more than sufficient for our needs, providing over an order of magnitude improvement in the internal kinematics of the association compared to previous works.

Of the 433 stars in our sample {\it Gaia} data are available for 258 (60\%). Figure~\ref{cmd} shows colour-magnitude diagrams for the three Sco-Cen subgroups with the sources lacking {\it Gaia} astrometry indicated. These sources are predominantly bright ($V < 6$~mag) and are likely missing from DR1 due to saturation \citep[the necessary processing required to accurately measure astrometry for such sources was not established in DR1,][]{lind16}. Additionally, sources with extremely red or blue colours, or those with high proper-motions ($>$3.5~arcsec/yr) are not included in DR1 \citep{brow16,aren17}, though this won't include any of our targets. Where {\it Gaia} data are unavailable we take parallaxes and proper motions from the revised {\it Hipparcos} catalogue \citep{vanl07}.

\begin{figure}
\begin{center}
\includegraphics[height=235pt, angle=270]{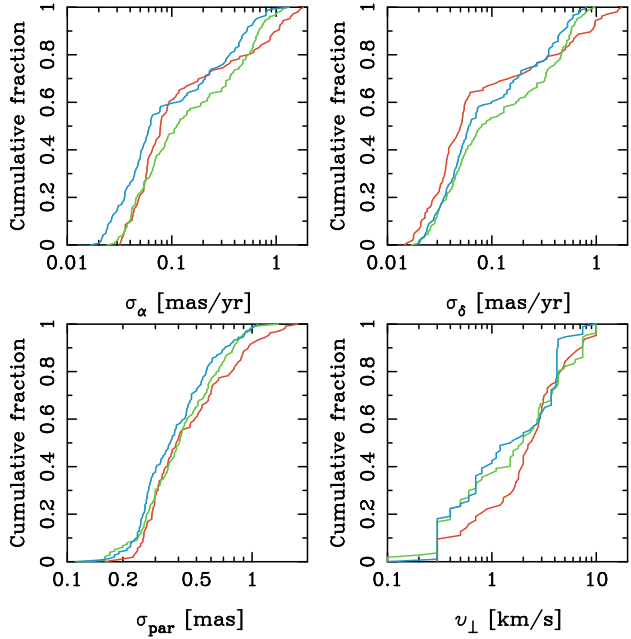}
\caption{Cumulative uncertainty distributions for the samples of stars in US (red), UCL (green), and LCC (blue) for parallax, proper motions, and RVs.}
\label{uncertainties}
\end{center}
\end{figure}

The precision of the astrometric data used is shown in Figure~\ref{uncertainties}. The median proper motion precision is 0.06--0.11~mas/yr ($\simeq$0.04--0.08 km/s at a representative distance of 150~pc) in $\alpha$ and 0.05-0.08~mas/yr ($\simeq$0.04--0.06 km/s) in $\delta$ for all the data for the three subgroups. When just {\it Gaia} data is considered the median proper motion precision is 0.04--0.06~mas/yr (0.03--0.06 km/s) in $\alpha$ and $\sim$0.04~mas/yr (0.03 km/s) in $\delta$. The median parallax uncertainty is $\sim$0.4~mas for all subsets of the data\footnote{{\it Gaia} DR1 provides considerably improved proper motions compared to {\it Hipparcos} because of the increased baseline of observations used, but the parallax precision is not significantly improved because the additional {\it Gaia} observations do not cover sufficient time to resolve the parallax motion of stars.}. Note that systematics in the astrometric solution used for DR1 result in an additional parallax systematic error of 0.3~mas that cannot be reduced by averaging parallaxes for groups of stars \citep{lind16}. Correlations between astrometric parameters for a single source can reach significant levels over large areas of the sky so in this work we make use of the full covariance matrix when calculating uncertainties on all astrometric parameters.

\subsection{Radial velocities}

Results from {\it Gaia}'s RV spectrometer are not included in DR1 and so we gathered RVs from the literature. We followed \citet{murp15} in the prioritisation of RVs from different samples, but chose not to exclude RVs from known binaries\footnote{Our knowledge of binarity in Sco-Cen is likely to be incomplete and so excluding only the known binaries will not eliminate the impact of binaries on our measurements, only reduce it by an unknown amount. Our approach instead is to model the impact binaries have (such as on the velocity dispersion) and exclude sources known to be binaries where this approach is not feasible.}. Our primary source of RVs was the Magellan/MIKE study of \citet{chen11}, from which 89 matches were found with our catalogue. Our next sources were the Pulkovo Compilation of Radial Velocities \citep[PCRV,][]{gont06}, from which 118 matches were found, and CRVAD-2 \citep{khar07}, from which 53 matches were found. For the latter catalogue we made sure to remove all sources included from the pre-publication PCRV list that are non-spectroscopic `astrometric' binaries not suitable for kinematic analysis. Finally we took 14 RVs from the Keck and Magellan spectra presented by \citet{dahm12}. In total this provided us with RVs for 274 out of 433 stars in our sample. The median RV uncertainty in each of the three subgroups is 2.3 (US), 2.0 (UCL), and 1.5~km/s (LCC).

\section{Distance and 3-dimensional structure}

In this section we calculate the distances to the centres of each subgroup, necessary for calculating the 3-dimensional bulk motions (Section~\ref{s-bulkmotions}) and for assessing the evidence for linear expansion (Section~\ref{s-expansion}). We also combine the parallaxes with spatial positions to fit 3-dimensional radial profiles to each subgroup, which are necessary for calculating their virial state (Section~\ref{r-virialstate}).

\subsection{Individual distances}
\label{s-individual}

Distances to individual sources and their uncertainties were calculated using the Bayesian inference method of \citet{bail15}, which overcomes the biases and limitations of traditional distance estimates derived from parallaxes. We do this for each star in our sample individually, employing a prior that converges asymptotically towards zero as the distance goes to infinity, $P(r) = r^2 \mathrm{exp} (-r / L)$, where $L = 1.35$~kpc is the scale length parameter, as recommended by \citet{bail15} and \citet{astr16}\footnote{The use of such a prior is not recommended when determining the distance to a cluster of stars and therefore we do not use it in Section~\ref{s-spatialmodel} when modelling the distances to each subgroup. The effect on our sample of stars of using such a prior over not using a prior is to increase individual distances by an average of $\sim$0.1~pc.}. To fully sample the posterior distribution we use the Markov Chain Monte Carlo (MCMC) ensemble sampler {\it emcee} \citep{fore13}, and use the mode of the corresponding posterior as our distance estimate and the 68\% confidence interval as the 1$\sigma$ uncertainty.

Due to the proximity of Sco-Cen and its large extent on the sky it is often necessary to consider the physical structure and spatial distribution in Galactic cartesian coordinates, $XYZ$, rather than in $\alpha$, $\delta$ and distance. We therefore calculate $XYZ$ coordinates for all sources using the same method as for the distance and accounting for the correlated uncertainties between $\alpha$, $\delta$, and $\varpi$. Note that the distance and $XYZ$ coordinate uncertainties are purely derived from the quoted random errors and do not include the systematic uncertainty.

\begin{figure}
\begin{center}
\includegraphics[height=240pt, angle=270]{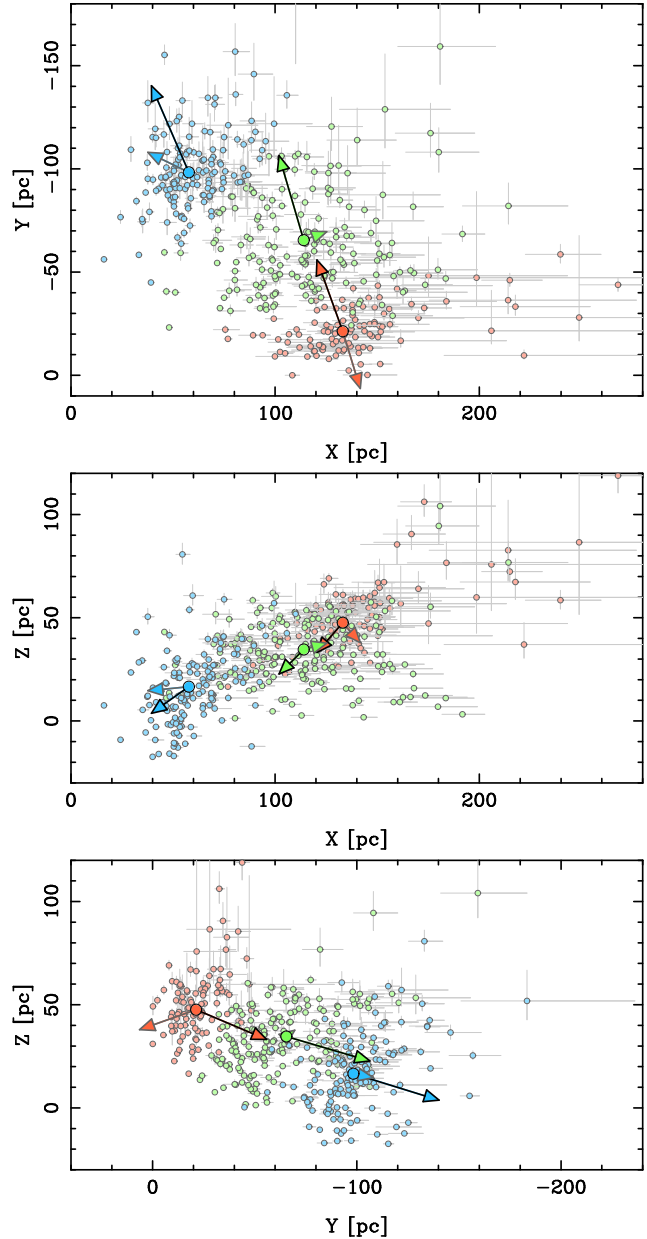}
\caption{3-dimensional spatial structure of Sco-Cen shown in the $X-Y$ (top), $X-Z$ (middle), and $Y-Z$ planes (bottom panel). The sources are coloured according to their subgroup members: red for US, green for UCL, and blue for LCC. 1$\sigma$ uncertainties on the positions of all sources are shown in grey. Also shown are the centres of each subgroup (large circles, see Section~\ref{s-spatialmodel}), the bulk motion of each subgroup over 2~Myrs (arrows with black outline, see Section~\ref{s-velocitymodel}), and the relative motion of each subgroup over 10~Myrs with the overall Sco-Cen bulk motion ($U = -7.0 \pm 0.2$, $V = -19.6 \pm 0.2$, $W = -6.1 \pm 0.1$~km/s) subtracted (arrow with grey outline).}
\label{bulkmotions}
\end{center}
\end{figure}

Figure~\ref{bulkmotions} shows the 3-dimensional spatial structure of Sco-Cen with the internal structure resolved in all three subgroups for the first time (\citealt{deze99} was unable to resolve the internal structure of any of the subgroups, while \citealt{debr99} was only able to resolve the internal structures of UCL and LCC with their improved secular parallaxes). The physical extents of the three subgroups overlap to varying degrees in the three dimensions, being most clearly separated in the $Y-Z$ plane, which is the closest equivalent to the observational $l-b$ plane. A number of spatial outliers from the three subgroups can be seen, especially in the $X-Y$ plane, though their uncertainties are sufficiently high that their membership of Sco-Cen cannot be rejected based solely on this.

\subsection{Modelling the subgroup distances and radial profiles using Bayesian inference}
\label{s-spatialmodel}

We use Bayesian inference to estimate the distance, size and radial profile of each subgroup. It is necessary to fit these quantities at the same time because the physical extent of each subgroup is a not-insignificant fraction of their distance and thus fits for the two will be correlated. 

The idea of Bayesian inference is to create a parameterised model that can reproduce the observations, and then compare that model (for different sets of parameters) to the observations in a probabilistic way. The aim of this process is to determine which of the various sets of parameters, $\boldsymbol{\theta}$, best explain the observations, $\boldsymbol{d}$. In Bayes's theorem this is known as the posterior distribution, $\boldsymbol{P(\theta | d)}$, and is given by

\begin{equation}
\boldsymbol{P(\theta | d)} =  \frac{ \boldsymbol{P(d | \theta)} \times \boldsymbol{P(\theta)} }{ \boldsymbol{P(d)} }
\end{equation}

\noindent where $\boldsymbol{P(d | \theta)}$ is the likelihood model, $\boldsymbol{P(\theta)}$ are the priors (which includes our a priori knowledge about the model parameters) and $\boldsymbol{P(d)}$ is a normalising constant.

We use Bayesian inference because it is better to project the model predictions into observational space, where the measurement uncertainties are defined, rather than deriving physical quantities and then trying to calculate appropriate physical uncertainties to use in comparison. This is particularly the case when the observations have highly correlated uncertainties, as in the case of DR1, or when parallaxes are involved \citep[see discussion in][]{bail15}.

\subsubsection{Method}

Our likelihood model constructs the 3D distribution of sources in each subgroup in Galactic cartesian coordinates ($XYZ$) and then reprojects the positions of individual stars into equatorial coordinates and distances, calculating parallaxes from the latter. We then add measurement uncertainties (including correlated uncertainties) to the parallax, $\alpha$ and $\delta$, randomly sampling these from the observed uncertainty distributions for our sources (a mixture of {\it Gaia} and {\it Hipparcos} uncertainties).

The structure of each subgroup is modelled using a 3D \citet[EFF,][]{elso87} radial profile, where the stellar surface density, $\Sigma(r)$, at a radius, $r$, is given by

\begin{equation}
\Sigma (r) = \Sigma_0 \left( 1 + \frac{r^2}{a^2} \right)^{-\gamma / 2}
\end{equation}

\noindent wherein $\Sigma_0$ is the central surface density, $a$ is the scale length, and $\gamma$ is the power-law index. The scale length is related to the half-mass, or effective, radius according to $r_{hl} = a \sqrt{4^{1/\gamma} - 1}$, while the power-law index can be used to calculate the quantity $\eta$ in the virial equation \citep[e.g.,][]{port10}.

We tried two variants of this model, the first with an isotropic radial profile (i.e., single values of $a$ and $\gamma$) and the second where the radial profiles in the three dimensions were allowed to vary. Our two models therefore have 5 (distance, central $\alpha$ and $\delta$, $a$, and $\gamma$) and 9 parameters (where $a$ and $\gamma$ become $a_X$, $a_Y$, $a_Z$, $\gamma_X$, $\gamma_Y$, and $\gamma_Z$).

We adopt liberal priors where possible, to minimise their impact on the fit. For the EFF parameters we use linear priors of 1--50~pc for the scale factors and 2--20 for the power-law indices. For the central subgroup distances we apply linear priors of 50--250~pc, while the priors on $\alpha$ and $\delta$ are left unconstrained.

To fully sample the posterior distribution function we use the MCMC ensemble sampler {\it emcee}, computing the likelihood model for each point in the parameter space and comparing the model to the observations using an unbinned maximum likelihood test. We found that the model fits would typically converge within 10,000 iterations, with a similar number of iterations necessary to fully sample the posterior distribution function. The posterior distribution function was typically found to follow a normal distribution, and thus the median value was used as the best fit, with the 68\% confidence interval used as the 1$\sigma$ uncertainty.

\begin{figure*}
\begin{center}
\includegraphics[height=510pt, angle=270]{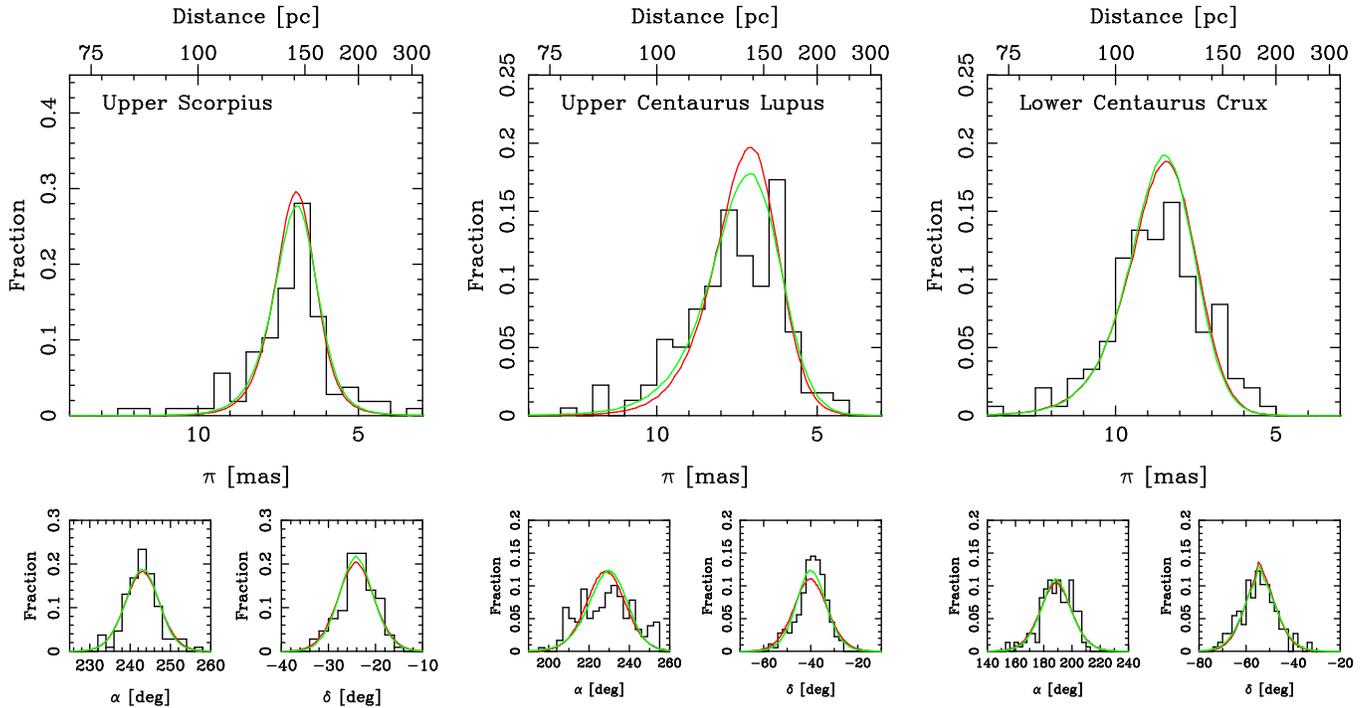}
\caption{Source distributions (black histogram) in parallax (top row), $\alpha$ and $\delta$ (bottom row) for the three subgroups of Sco-Cen compared to the EFF model fit results for 1D (red) and 3D (green) models. The model fit distributions were calculated by sampling the best-fitting forward models 1,000,000 times and plotting the resulting probability distribution functions.}
\label{distance_fits}
\end{center}
\end{figure*}

\begin{table}
\caption{Structural and kinematic properties of the three Sco-Cen subgroups determined in this work. The structure parameters are all determined from the 1D EFF model, but these are generally in close agreement with the 3D EFF model. The distance uncertainties do not take into account the 0.3~mas systematic parallax uncertainty present in DR1. The velocity dispersion fitting results are taken from the model using a binary fraction of 100\%.}
\label{structure} 
\begin{tabular}{lccc}
\hline
Subgroup				& US						& UCL					& LCC				\\
\hline
Distance [pc]			& $143.0^{+0.3}_{-0.4}$		& $135.9^{+0.5}_{-0.4}$		& $115.2^{+0.3}_{-0.3}$	\\
$\alpha_0$ [deg]		& $243.02^{+0.15}_{-0.10}$	& $228.36^{+0.27}_{-0.24}$	& $188.57^{+0.23}_{-0.19}$ \\
$\delta_0$ [deg]		& $-24.19^{+0.08}_{-0.18}$	& $-40.50^{+0.13}_{-0.13}$	& $-54.53^{+0.20}_{-0.06}$ \\
$X_0$ [pc]			& $133.1^{+0.4}_{-0.3}$		& $114.0^{+0.5}_{-0.5}$		& $57.7^{+0.3}_{-0.3}$	\\
$Y_0$ [pc]			& $-21.3^{+0.2}_{-0.4}$		& $-65.5^{+0.4}_{-0.5}$		& $-98.3^{+0.3}_{-0.3}$	\\
$Z_0$ [pc]			& $47.6^{+0.3}_{-0.3}$		& $34.6^{+0.4}_{-0.3}$		& $16.6^{+0.3}_{-0.2}$	\\
$a$ [pc]				& $35.8^{+1.9}_{-1.3}$		& $69.8^{+3.3}_{-3.2}$		& $50.1^{+4.0}_{-3.2}$	\\
$\gamma$			& $14.5^{+1.4}_{-1.0}$		& $17.6^{+1.3}_{-1.3}$		& $15.2^{+1.9}_{-1.5}$	\\
$r_{hl}$ [pc]			& $11.4^{+0.2}_{-0.2}$		& $20.0^{+0.3}_{-0.2}$		& $15.5^{+0.3}_{-0.2}$	\\
$U_0$ [km/s]			& $-6.16_{-0.13}^{+0.14}$		& $-5.90_{-0.12}^{+0.17}$		& $-8.96_{-0.24}^{+0.14}$	\\
$V_0$ [km/s]			& $-16.89_{-0.10}^{+0.08}$	& $-20.00_{-0.10}^{+0.12}$	& $-20.55_{-0.13}^{+0.16}$ \\
$W_0$ [km/s]			& $-7.05_{-0.08}^{+0.09}$		& $-5.80_{-0.09}^{+0.09}$		& $-6.29_{-0.11}^{+0.12}$	\\
$\sigma_U$ [km/s]		& $1.63_{-0.20}^{+0.20}$		& $1.96_{-0.12}^{+0.13}$		& $1.89_{-0.13}^{+0.21}$	\\
$\sigma_V$ [km/s]		& $1.14_{-0.14}^{+0.13}$		& $0.73_{-0.19}^{+0.15}$		& $0.90_{-0.30}^{+0.50}$	\\
$\sigma_W$ [km/s]		& $2.51_{-0.09}^{+0.11}$		& $1.27_{-0.14}^{+0.11}$		& $0.51_{-0.18}^{+0.25}$	\\
$\sigma_{3D}$ [km/s]	& $3.20_{-0.20}^{0.22}$		& $2.45_{-0.20}^{+0.20}$		& $2.15_{-0.24}^{+0.47}$	\\
$\alpha_{CP}$ [deg]		& $116.22_{-9.46}^{+10.70}$	& $100.88_{-1.52}^{+1.69}$	& $95.32_{-1.31}^{+1.29}$ \\
$\delta_{CP}$ [deg]		& $-55.29_{-4.56}^{+5.37}$	& $-35.88_{-1.61}^{+1.63}$	& $-28.68_{-1.48}^{+1.53}$ \\
\hline
\end{tabular}
\end{table}

\subsubsection{Results}

The results of the model fits are provided in Table~\ref{structure} and illustrated in Figure~\ref{distance_fits}. Both the 1D and 3D EFF models show reasonable fits to the data for US and LCC, though the fits for UCL are not as good, most likely because this subgroup is clearly less centrally concentrated, particularly in $\alpha$, and therefore an EFF model is probably not appropriate. However, our purpose here is to obtain a reliable estimate of the distance to each subgroup and their physical sizes and so these results should be sufficient.

The central coordinates of each subgroup in all three dimensions show good agreement between the 1D and 3D model fit results, with all values within 1$\sigma$ of their corresponding value. Combining the systematic 0.3~mas parallax uncertainty with the uncertainties already calculated leads to distances for the three subgroups of $143 \pm 6$, $136 \pm 5$, and $115 \pm 4$~pc (though if the systematic uncertainties have been over-estimated then these uncertainties will also be over-estimated). Our values are consistent with the mean distances found by \citet{deze99} of $145 \pm 2$, $140 \pm 2$, and $118 \pm 2$, with our distances slightly smaller than theirs, most likely because of the changes to the membership of Sco-Cen performed by \citet{rizz11}, and with slightly larger uncertainties \citep[because of the systematic and spatially-correlated 0.3~mas parallax uncertainty in DR1 that is not reduced by averaging,][]{brow16}. Note that because the physical extent of each subgroup is larger than the uncertainty on the central distances calculated here these distances should only be considered as representative distances for each subgroup and not as distances for the individual stars.

The power-law indices of the 1D EFF profiles are found to be 14.5, 17.6 and 15.2 for US, UCL and LCC respectively, all of which agree with their 3D EFF indices and with each other to within $\sim$1$\sigma$. The 1D EFF scale lengths are 35.8, 69.8 and 50.1~pc for US, UCL and LCC respectively. The scale lengths show good agreement between the 1D and 3D models. Both US and LCC appear close to spherical, with similar scale lengths in all three dimensions that differ by less than 10\% from each other and from the 1D scale length.

UCL is the least spherical of the subgroups, elongated in the $X$ direction relative to the $Y$ and $Z$ directions with scale lengths of 74, 57, and 48~pc, respectively. If the subgroups of Sco-Cen do not represent distinct episodes of star formation as some recent studies have suggested \citep[e.g.,][]{rizz11,peca16} but are actually made up of smaller substructures, this would explain the non-spherical structure of UCL. The true 3D structure is likely to be much more complex, and also not aligned with the $XYZ$ coordinate system. Since the differences between the 1D and 3D scale lengths is not dramatic, and because we only require a simple estimate of the size of our subgroup, we will use the 1D scale lengths throughout the rest of this work.

\section{Bulk Motions and Velocity Dispersions}
\label{s-bulkmotions}

Here we use proper motions and RVs to measure the basic kinematic quantities of each subgroup, their bulk motions, velocity dispersions, and convergence points. These are necessary for a full understanding of the dynamics of the association, to study their relative motions, assess their virial states, and to test whether the subgroups are expanding.

\subsection{Forward modelling the velocity dispersion}
\label{s-velocitymodel}

To measure the velocity dispersion and bulk motion of each subgroup we use Bayesian inference, as in Section~\ref{s-spatialmodel}. We model these quantities in the Galactic cartesian system $XYZ$, with velocities $UVW$, and then convert them to the observational frame. We adopt the classical definition of this coordinate system with $X$ directed towards the Galactic centre and increasing in that direction, $Y$ positive in the direction of Galactic rotation, and $Z$ positive towards the north Galactic pole.

\subsubsection{Method}

Our forward model begins by modelling a population of stars with 3D coordinates, $XYZ$, sampled from the observed distribution of these coordinates for each subgroup, as calculated in Section~\ref{s-individual}. The 3D velocities of each star are randomly sampled from a trivariate Gaussian velocity distribution and then reprojected into the observational frame to obtain modelled proper motions and RVs for each star.

Because we are modelling RVs it is necessary to model the contribution that binary orbital motions make to the observed velocity distribution. To do this we simulate a population of randomly-aligned binaries in a manner similar to that of \citet{oden02} and \citet{cott12}, randomly selecting primary and secondary masses, semi-major axes, and eccentricities for each binary. The primary masses were selected from a standard $\alpha = 2.3$ initial mass function \citep{krou01} in the mass range of 1.4--17.5~M$_\odot$ (appropriate for the A and B-type stars that constitute our sample) using the equations of \citet{masc13}.

\begin{figure*}
\begin{center}
\includegraphics[height=510pt, angle=270]{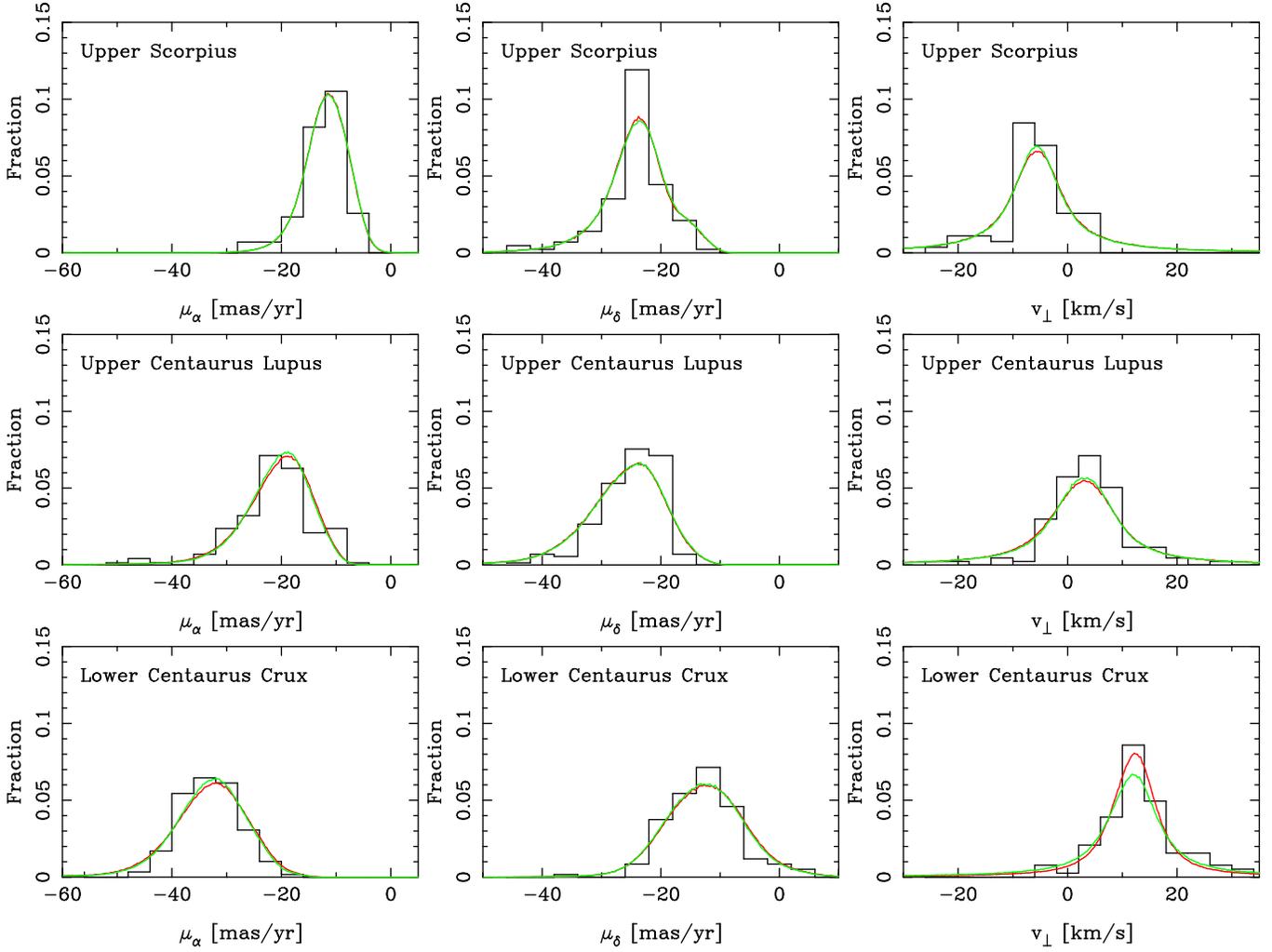}
\caption{Source distributions (black histograms) for proper motions (left and centre) and RVs (right) for the three subgroups, compared to the best-fitting velocity dispersion model fit results for binary fractions of 100\% (green) and 70\% (red). The model fit distributions were calculated by sampling the best-fitting forward models 1,000,000 times and plotting the resulting probability distribution functions.}
\label{velocity_dispersions}
\end{center}
\end{figure*}

We use the orbital properties of intermediate-mass stars in Sco-Cen determined by \citet{kouw07}, who performed an extensive study of the visual, spectroscopic, and astrometric binaries in the association. They inferred a binary fraction of 100\%, a mass ratio distribution $f_q(q) \propto q^{-0.4}$, and a semi-major axis distribution of the form $f_a(a) \propto a^{-1}$ (in the range $5 \geq a / R_\odot \geq 5 \times 10^6$). The authors were unable to sufficiently constrain the eccentricity distribution from the available observations and so we adopt a flat distribution in the range of 0--1, which \citet{kouw07} found to be consistent with their observations. We then calculate instantaneous velocity offsets for the primary star relative to the centre of mass of the system for a random phase within the binary's orbit. Following this method we measure the velocity offsets for $10^6$ random binaries, providing us with a distribution from which we can sample. We don't consider triple systems because their properties are poorly constrained \citep{duch13} and are typically hierarchical, meaning that the third star is usually on a wide, long-period orbit that does not have a large velocity.

Stars that are modelled as being in a binary system have velocity offsets added to their RV, randomly sampled from our velocity offset distribution. Finally we add measurement errors for the proper motions and RVs for each star, randomly sampling these from the empirical uncertainty distribution for the stars in each subgroup.

This model has 6 free parameters, the central velocity ($U_0$, $V_0$, $W_0$) and velocity dispersion ($\sigma_U$, $\sigma_V$, $\sigma_W$) in each dimension, for which we adopt flat and wide priors of $-100 \leq v_0 \mathrm{[km/s]} \leq +100$ for the central velocities and $0 \leq \sigma \mathrm{[km/s]} \leq 100$ for the velocity dispersions. As in Section~\ref{s-spatialmodel} we sampled the posterior distribution function using an MCMC ensemble sampler, computed the likelihood model for each point in parameter space, and compared the model to the observations using an unbinned maximum likelihood test. The posterior distribution functions followed a broadly normal distribution, albeit with wide tails, and so the median value was used as the best fit and the 68\% confidence interval used for the 1$\sigma$ uncertainty.

\subsubsection{Results}

\begin{figure*}
\begin{center}
\includegraphics[width=510pt]{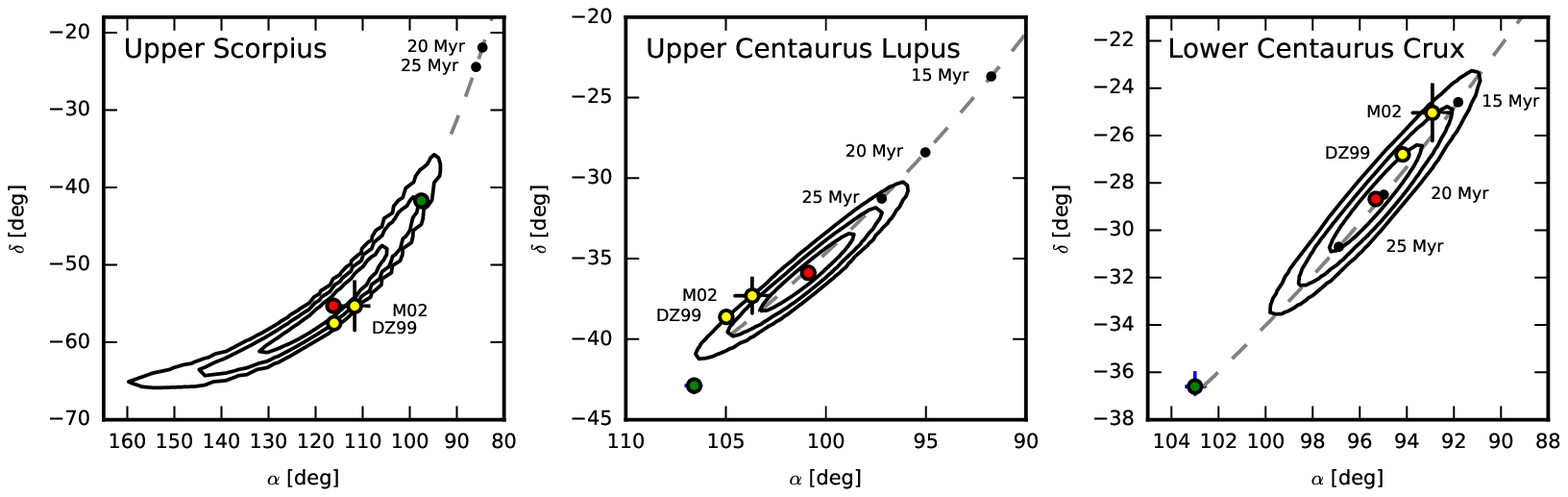}
\caption{Convergent points for the three subgroups of Sco-Cen determined using the \citet{jone71} method (red circle and contours showing 1, 2, and 3$\sigma$ uncertainties) and from the bulk motions (green circles with 1$\sigma$ error bars typically equal to or smaller than the size of the symbols). Yellow circles show previously published convergent points from \citet[][DZ99]{deze99} and \citet[][M02]{mads02}, with 1$\sigma$ error bars shown for the latter (uncertainties for the former are expected to be several degrees along the great circle connecting the convergent point with the subgroup). Dashed lines with black markers show the predicted convergent points derived from the mean space motions at a range of expansion ages.}
\label{convergent_point}
\end{center}
\end{figure*}

The results are provided in Table~\ref{structure} and illustrated in Figure~\ref{velocity_dispersions}, with the central velocities, or bulk motions, also shown in Figure~\ref{bulkmotions}. The bulk motions are different from those presented by \citet{deze99}, particularly in $U$ and $V$, implying that the difference is probably due to the increased availability and quality of RVs for these stars \citep[though the change in Sco-Cen membership by][will have had a small effect]{rizz11}. Our bulk motions are in better agreement with those presented by \citet{chen11}, with most agreeing within 1--2$\sigma$, and the remaining differences probably due to the change in membership.

The best-fitting velocity dispersions vary from 0.5 -- 2.5 km/s between the three subgroups of Sco-Cen and along the three axes. None of the subgroups have isotropic velocity dispersions, each having one dimension with a significantly larger dispersion than the other two dimensions. For US the velocity dispersion is largest in $W$, while for UCL and LCC it is largest in $U$. This is unlikely to be due to the influence of incorrectly-modelled binary systems since RVs contribute most to $U$ for US and UCL, and $V$ for LCC.

Our velocity dispersions are mostly consistent with previous estimates in the literature. \citet{debr99} found that the 1D velocity dispersions for all the subgroups are $\leq$~1.0-1.5~km/s, while \citet{mads02} measured velocity dispersions of 1.1--1.3~km/s for the subgroups. Our weighted mean 1D velocity dispersions for the three subgroups are $1.86 \pm 0.21$ (US), $1.38 \pm 0.21$ (UCL) and $1.21 \pm 0.28$ km/s (LCC), for which only the US velocity dispersion is inconsistent with previous findings (to approximately 1.9 and 2.8$\sigma$ for the two studies cited previously). The disagreement for US can be attributed to a combination of the revised membership of Sco-Cen\footnote{We repeated these fits using the \citet{deze99} membership list and find that the change in sample to \citet{rizz11} has caused the velocity dispersion to increase by $\sim$10\%.} and our inclusion of RVs to reveal the full 3D velocity dispersions (because of the anisotropy this causes a large change in the mean 1D velocity dispersion).

\citet{kouw07} find that the observations of Sco-Cen are best fit with a binary fraction of 100\%, though slightly smaller fractions are possible. We explored using binary fractions of 70--100\% and found that lower fractions lead to larger velocity dispersions in each dimension, with the largest difference seen along the axis contributing most to the RVs ($U$ for US and UCL, $V$ for LCC). The velocity dispersion increases by $\sim$10\% along that axis for each 10\% by which the binary fraction is reduced. Figure~\ref{velocity_dispersions} shows the model fit results for a 70\% binary fraction, with little difference in the fit quality compared to that for 100\% binaries.

\subsection{Convergent point motion}
\label{s-convergent}

Nearby groups of stars with a common space motion have proper motions that appear to `converge' towards a single point on the sky. This `convergent point' can be used to establish membership of these groups, measure their mean distances, and test models for their expansion (Section~\ref{s-expanding_cp}). In this section we measure the convergent points of the three subgroups using both the refurbished \citet{jone71} method \citep{debr99b} and directly from the 3D bulk motions.

\subsubsection{Method}

Jones's method iterates over a hemispherical grid of points to find the convergent point coordinates that minimises the proper motions perpendicular to the great circle joining each star to the convergent point. The best fitting convergent point is that which minimises the quantity $\sum t^2_\perp$, where $t_\perp = \mu_\perp / \sigma_\perp$ is the ratio of the perpendicular proper motion to its uncertainty, and the sum is over all stars in the group. This method was refined by \citet{debr99b} who changed this quantity to $t_\perp = \mu_\perp / \sqrt{\sigma^2_\perp + \sigma^2_v}$, recognising that moving groups and associations have an intrinsic, one-dimensional velocity dispersion, $\sigma_v$, that prevents all their proper motions from being directed exactly towards the convergent point and should therefore be accounted for. We adopt this approach, using an MCMC method to search for the convergent point, instead of a grid-based technique.

Since the $t_\perp^2$ value can be treated as the classic $\chi^2$ statistic it can be used to assess the reliability of a fit using the probability, $\epsilon$, that $t_\perp^2$ will exceed the observed value of $t_\perp^2$ by chance even for a correct model. Therefore, if the value of $t_\perp^2$ is too high the probability $\epsilon$ will be below the threshold value, which \citet{debr99b} recommend setting to a value of 0.954 (giving a $\sim$5\% probability of falsely rejecting the null hypothesis). If $\epsilon$ is below this value then the star with the highest value of $| t_\perp |$ is removed from the sample and the fitting process restarted. This process continues until a set of stars returns a convergent point with a sufficiently low $t_\perp^2$ value.

\subsubsection{Results}

Convergent points were found for all three subgroups of Sco-Cen using the Jones method and the mean 1D velocity dispersions determined in Section~\ref{s-velocitymodel}\footnote{Since the velocity dispersions are not isotropic one should ideally use the component of the velocity ellipsoid perpendicular to the great circle connecting the subgroup with the convergent point. However this calculation would be complex to perform for every convergent point considered, and since the exact velocity dispersion used does not significantly affect the final convergent point calculated we have instead used the mean 1D velocity dispersion.}. For US a convergent point for the entire subgroup was found with a $\chi^2$ probability of 53\%, rising to 95.4\% after removing the 20 stars with the highest $| t_\perp |$ values. The two convergent points vary by $\sim$0.05~deg, significantly less than their uncertainties. Higher velocity dispersions of 2.0 or 2.5~km/s resulted in convergence after rejecting only 6 or 0 stars, but these do not show significant variations from those already determined, so we choose the convergent point calculated using the mean 1D velocity dispersion of 1.86~km/s. Figure~\ref{convergent_point} shows this convergent point compared to those determined previously by \citet{deze99} and \citet{mads02} using just proper motions, which agree well despite the change in membership of the subgroup, and the convergent point determined from the mean space motion, which agrees to within 2$\sigma$.

For UCL a convergent point for the entire subgroup was found with a $\chi^2$ probability of 98\% without discarding any stars, and is therefore much better constrained than that of US. For LCC a convergent point for the entire subgroup was found with a $\chi^2$ probability of 20\%, rising to 95.4\% after removing the 10 stars with the highest $| t_\perp |$ values. A higher velocity dispersion of 1.5 km/s would result in an immediate fit with a $\chi^2$ probability of $>$99\% for all the stars in the subgroup. However, these three convergent points all agree within $\sim$0.05~deg, significantly less than the uncertainty, so the fit with the mean velocity dispersion of 1.21~km/s was used. For both UCL and LCC our convergent points agree with previous estimates in the literature to within 1--2$\sigma$, with any differences most likely due to changes in the subgroup membership. The agreement between the convergent points calculated using only proper motions and those calculated using the mean space motions are not as good however, disagreeing by more than 3$\sigma$. These disagreements suggest that some degree of expansion (or contraction) may be present.

\section{Expansion of the association}
\label{s-expansion}

In this section we explore whether the subgroups are in the process of expanding, and if so whether they are expanding from a single point. Due to the proximity of Sco-Cen we cannot use proper motions alone to study the expansion of the association on the plane of the sky since a radial motion of the association towards (or away from) the observer will cause an apparent expansion (contraction) of the association, even when none is actually present \citep{blaa64b}. We must therefore either combine the proper motions with RVs, which we can do using many different methods, or use the bulk radial motion to correct this {\it virtual expansion / contraction}.

\subsection{Linear expansion}
\label{s-linearexpansion}

\begin{figure}
\begin{center}
\includegraphics[height=240pt,angle=270]{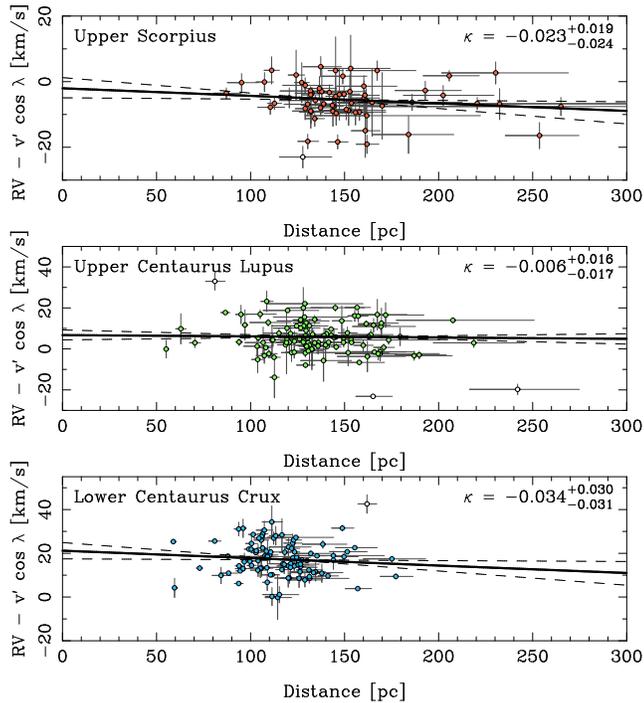}
\caption{Distance to sources in each subgroup of Sco-Cen plotted against the difference between the observed RV and that predicted from the moving group method ($v^\prime \, \mathrm{cos} \, \lambda$). The error bars for individual sources show the $16 - 84$\% confidence intervals, calculated from Monte Carlo simulations. The solid lines show the best fit linear relationship between the plotted quantities, with 1$\sigma$ uncertainties shown with dashed lines. The best fitting slopes, $\kappa$, and uncertainties are noted in each panel. If the subgroups are not expanding we would expect a slope of zero, while if they expanding we would expect a positive slope. 3$\sigma$ outliers on the ordinate, shown with non-coloured symbols, were excluded from the fit.}
\label{Blaauw_expansion}
\end{center}
\end{figure}

If an association is expanding then the observed RVs of individual stars will vary from those predicted from the moving group method for parallel motion by an amount that varies with distance. Stars on the near-side (far-side) of the association will have smaller (larger) RVs than would be predicted by the moving group method as these stars are moving away from the centre of the association and towards (away from) us.

In the \citet{blaa64b} linear expansion model individual RVs, $v_\mathrm{rad}$, are predicted to follow the relation

\begin{equation}
v_\mathrm{rad} = v^\prime \, \mathrm{cos} \, \lambda^\prime + \kappa d + K
\label{blaauw1}
\end{equation}

\noindent where $\lambda^\prime$ is the angular separation between star and convergent point (calculated using just the proper motions, see Section~\ref{s-convergent}), $d$ is the distance to the star in pc, $\kappa$ is an expansion term, and $K$ is a zero point (that can represent gravitational redshift or convective blueshift). $v^\prime$ is the barycentric velocity of a star in the association, given by

\begin{equation}
v^\prime = \frac{ \mu_v A }{ \varpi \, \mathrm{sin} \, \lambda^\prime }
\end{equation}

\noindent where $\mu_v$ is the proper motion in the direction of the convergent point (in mas/yr), $\varpi$ is the parallax (in mas), and $A = 4.74$~km~yr~s$^{-1}$ is the astronomical unit in km~yr~s$^{-1}$.

An expanding association will have $\kappa > 0$, from which an expansion age, $\tau = (\gamma \kappa)^{-1}$, (in Myrs) can be derived ($\gamma$ is a conversion factor of 1.0227~pc~Myr$^{-1}$~km$^{-1}$~s).

Figure~\ref{Blaauw_expansion} shows the difference between the observed RVs and those predicted from the moving group method plotted against the distance to each star. The solid lines show the best-fitting linear relationships between these quantities with 3$\sigma$ outliers on the ordinate excluded. Fits were obtained using the MCMC code {\it emcee}, accounting for the distance uncertainties by randomly varying the distances according to their uncertainties. Given the possibility that the uncertainties on the ordinate have been underestimated, these fits were obtained by introducing a factor, $f$, by which the uncertainties have been underestimated \citep[see][]{hogg10} and then marginalising over this parameter to obtain the uncertainties on the resulting fit parameters.

The best-fitting slopes for the three subgroups are $-0.023^{+0.019}_{-0.024}$ (US), $-0.006^{+0.016}_{-0.017}$ (UCL), and $-0.034^{+0.030}_{-0.031}$ km~s$^{-1}$~pc$^{-1}$ (LCC). All three subgroups exhibit marginally negative slopes that are consistent with a slope of zero within a little over 1$\sigma$. The results for US are consistent with the fit of $\kappa = -0.01 \pm 0.04$~km~s$^{-1}$~pc$^{-1}$ found by \citet{peca12}. Based on these slopes and their confidence intervals we can rule out expansion ages equal to or smaller than the median ages of 10, 16 and 15~Myr found by \citet{peca16} with confidences of approximately 7, 4, and 3$\sigma$ for the three subgroups respectively. The fits were repeated with known binaries excluded (10, 1, and 2 sources in the three subgroups), with no significant difference.

\subsection{3D linear expansion}
\label{s-3Dexpansion}

\begin{figure*}
\begin{center}
\includegraphics[height=510pt,angle=270]{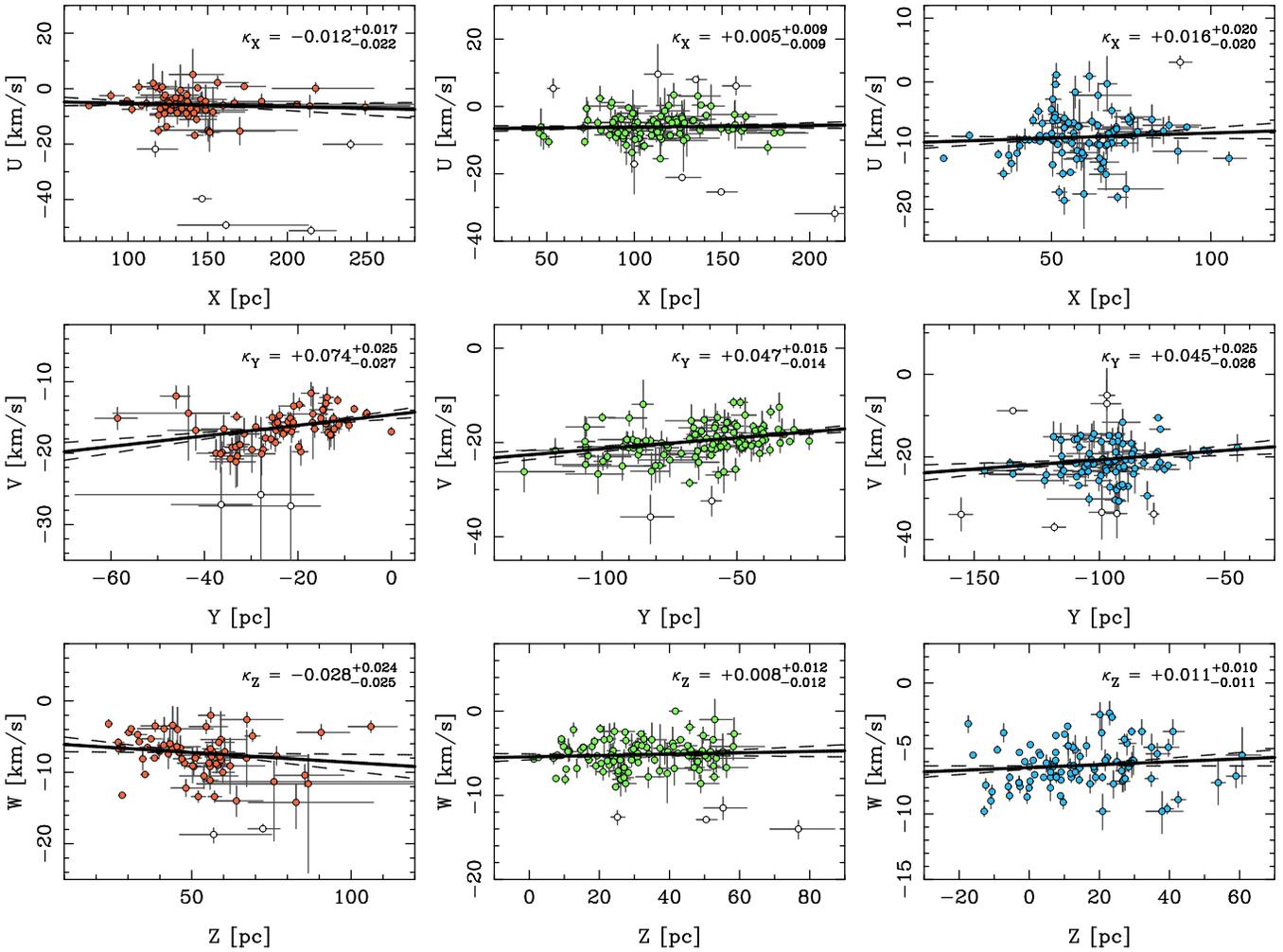}
\caption{Positions versus velocities along the three Galactic Cartesian axes $XYZ$ for each of the three subgroups of Sco-Cen (US on the left, UCL in the centre and LCC on the right). 1$\sigma$ error bars are shown for all sources. The solid lines show the best fit linear relationships between the plotted quantities, with 1$\sigma$ uncertainties shown with dashed lines. The best fitting slopes, $\kappa$, and uncertainties are noted in each panel. If the subgroups are not expanding we would expect slopes of zero, while if they are expanding we would expect a positive slope (a negative slope implies contraction along that axis). 3$\sigma$ outliers on the ordinate, shown with non-coloured symbols, were excluded from the fits.}
\label{UVW_vs_XYZ}
\end{center}
\end{figure*}

Using the three-dimensional positions and velocities (for the stars with RVs) we can search for evidence of expansion in each of the three axes of the Galactic Cartesian coordinate system by plotting velocity versus position and searching for evidence of a positive slope between the two quantities that would provide evidence for expansion along that axis.

Figure~\ref{UVW_vs_XYZ} shows the velocities $UVW$ plotted against the positional coordinates $XYZ$ for each of the three subgroups. We fit linear relationships to these quantities using MCMC, as above (with 3$\sigma$ outliers on the ordinate excluded), to estimate the slopes, $\kappa_X$, $\kappa_Y$, $\kappa_Z$, in the three dimensions. For US the slopes are $(\kappa_X, \kappa_Y, \kappa_Z) = (-0.01 \pm 0.02, 0.07 \pm 0.03, -0.03 \pm 0.03)$ km~s$^{-1}$~pc$^{-1}$, for UCL they are $(0.01 \pm 0.01, 0.05 \pm 0.02, 0.01 \pm 0.01)$ km~s$^{-1}$~pc$^{-1}$ and for LCC they are $(0.02 \pm 0.02, 0.05 \pm 0.03, 0.01 \pm 0.01)$ km~s$^{-1}$~pc$^{-1}$.

In the $X$ and $Z$ dimensions all three subgroups show either marginally negative or positive slopes that are consistent with a slope of zero within 1$\sigma$, implying no evidence for either expansion or contraction. Interestingly however all three subgroups show significant evidence for expansion along the $Y$ axis, with significances of approximately 3$\sigma$ for US and UCL and 2$\sigma$ for LCC. The fits were repeated with known binaries removed, as above, but there was no significant difference in the results.

\subsection{Expanding versus non-expanding convergent points}
\label{s-expanding_cp}

\begin{figure*}
\begin{center}
\includegraphics[height=510pt,angle=270]{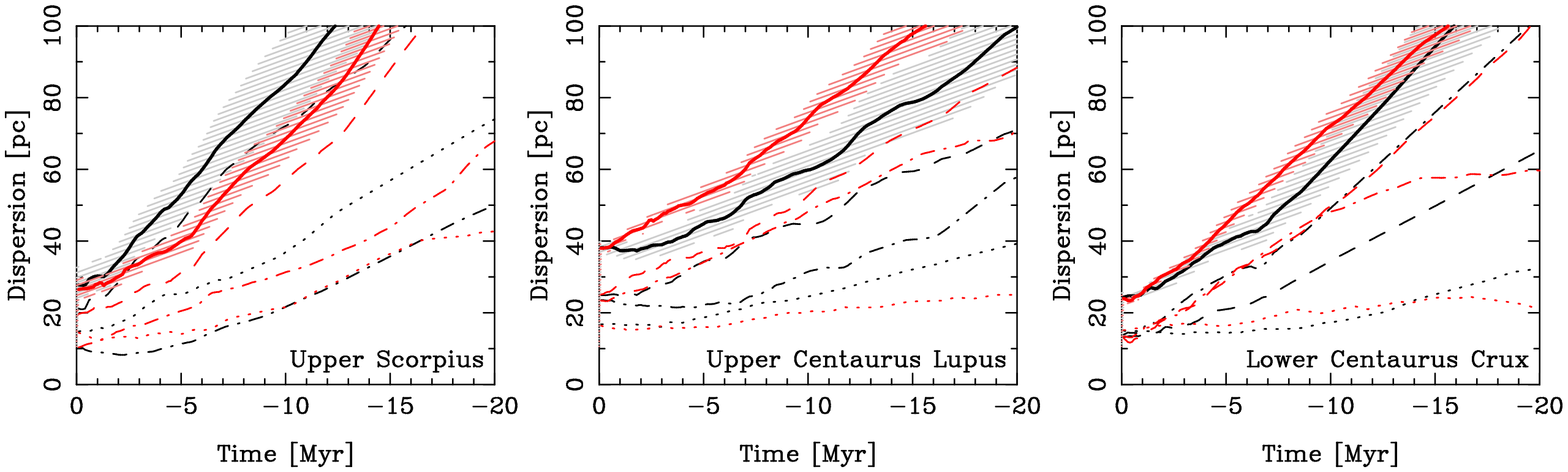}
\caption{1$\sigma$ dispersions (16th to 84th percentiles used to minimise the impact of outliers) of the size of each subgroup in each axis of the $XYZ$ coordinate system ($\sigma_X$ shown with a dashed line, $\sigma_Y$ with a dotted line, and $\sigma_Z$ with a dash-dotted line) as well as the quadrature sum of all three dimensions (solid line). Black lines show the results of traceback analysis performed with linear trajectories and the red lines show the analysis performed using the epicycle orbit approximation. Shaded areas around the 3D sum lines show the 90\% confidence interval in the quadrature sum dispersion of each subgroup determined from Monte Carlo simulations exploring the impact of uncertainties in the velocities.}
\label{closest_approach}
\end{center}
\end{figure*}

Our calculation of the convergent points that used only proper motions were based on the assumption that the associations were not expanding. However, a group of stars that has a small linear expansion will appear to converge to a point further away (i.e. with higher $\lambda$) than for a group of stars without expansion. This could explain the disagreement between the convergent points calculated using just proper motions and those calculated from the centroid space motion. To test whether the subgroups are expanding we add linear expansion to the centroid space motions (adding a term $\kappa d$ to the RVs, as per equation~\ref{blaauw1}) and compare these with the proper motion convergent points.

Figure~\ref{convergent_point} shows the centroid space motion convergent points for various expansion ages in the range 0--100~Myr. For US adding expansion actually worsens the agreement between the two convergent points, which argues against this subgroup exhibiting significant levels of expansion. For UCL small levels of expansion lead to an improved agreement between the convergent points for expansion ages $\gtrsim$20~Myr, though this is larger than its estimated age of $15 \pm 3$~Myr \citep{peca16}. For LCC an expansion age of 18--25~Myr leads to a good agreement between the two convergent points, which is in reasonable agreement with the estimated age of $16 \pm 2$~Myr from \citet{peca16}. Based on this analysis both US and UCL appear to be inconsistent with being in a state of expansion, whilst LCC is consistent with being an expanding association.

\subsection{Tracing back individual stellar motions}
\label{s-traceback}

An alternative method to assess the expansion of a group of stars is to trace back their individual motions in 3D and quantify the spatial extent at each time-step. For OB associations a simple traceback is generally valid because the densities are not high enough for there to be a significant number of close encounters and scatterings that would invalidate this approach. We do this considering both linear trajectories, i.e., $X(t) = X_0 + \gamma U t$ for the $X$/$U$ dimension (where the time, $t$, is in Myr and $\gamma = 1.022$~s~pc~km$^{-1}$~Myr$^{-1}$), and using an epicycle approximation.

Figure~\ref{closest_approach} shows the 1$\sigma$ dispersions (16th to 84th percentiles) in $X$, $Y$, $Z$, and in 3D using linear trajectories. For all three subgroups the dispersion in $X$ steadily increases as one goes further back into the past. The dispersion in $Y$ is similar for US, but for UCL and LCC it is approximately constant over the past $\sim$5 and $\sim$10~Myr, respectively, before showing signs of increase. The dispersion in $Z$ varies the most between subgroups. For LCC it is steadily increasing, for UCL it is either unchanged or slightly decreasing over the last $\sim$5~Myr and then increases, while for US it shows evidence for a slight decrease over the last $\sim$3~Myr before increasing. The 3D dispersions steadily increase as one goes back into the past, with only UCL showing evidence for a minimum not at the present day.

For a more accurate traceback of the stellar motions we use the epicycle approximation, employing the orbital equations from \citet{fuch06}. We use the Oort A and B constants from \citet{feas97}, the local disc density from \citet{holm04}, the local standard of rest velocity from \citet{scho10} and a solar $Z$ distance above the Galactic plane of 17~pc \citep{kari17}. Figure~\ref{closest_approach} shows the 1$\sigma$ spatial dispersions using the epicycle approximation. The results do not significantly differ from those calculated using linear trajectories, with all subgroups and all dimensions showing a broad increase in dispersion as we look further back in time. The largest increases are universally in the $X$ dispersion, with the smallest increases in the $Y$ dispersion (and in fact for UCL and LCC the dispersions in $Y$ are fairly steady over the lifetimes of the subgroups), and significant increases in the 3D dispersions.

Uncertainties in velocity can cause an expanding subgroup to appear larger in the past as the measured velocities differ from a true expansion pattern. To test the significance of this effect we performed a Monte Carlo simulation by randomly varying the measured velocities in accordance with their uncertainties and then recalculating the traceback motions. We repeated this 10,000 times and in Figure~\ref{closest_approach} we show the 90\% confidence interval around the 3D dispersions. The uncertainties in the size of each subgroup in the past are notably smaller than their actual size, suggesting that this effect is not due to measurement error.

To conclude, using either linear trajectories or an epicycle approximation and defining the size of the subgroups from their 1$\sigma$ spatial dispersions there is no evidence that the three subgroups are expanding or that they have ever had a more compact configuration in the past. It is worth noting that \citet{brow97} explored the validity of similar methods to this by producing synthetic proper motion observations of expanding OB associations and found that the smallest size of the OB association predicted from a traceback analysis using proper motions alone was always overestimated. The improvements in proper motion precision from {\it Gaia} since this work, as well as the increased availability and precision of RVs that allow this analysis to be performed in 3D rather than in the plane of the sky (and thus overcome the effects of radial streaming), do resolve many of the issues raised by \citet{brow97} however.

\begin{figure*}
\begin{center}
\includegraphics[width=510pt]{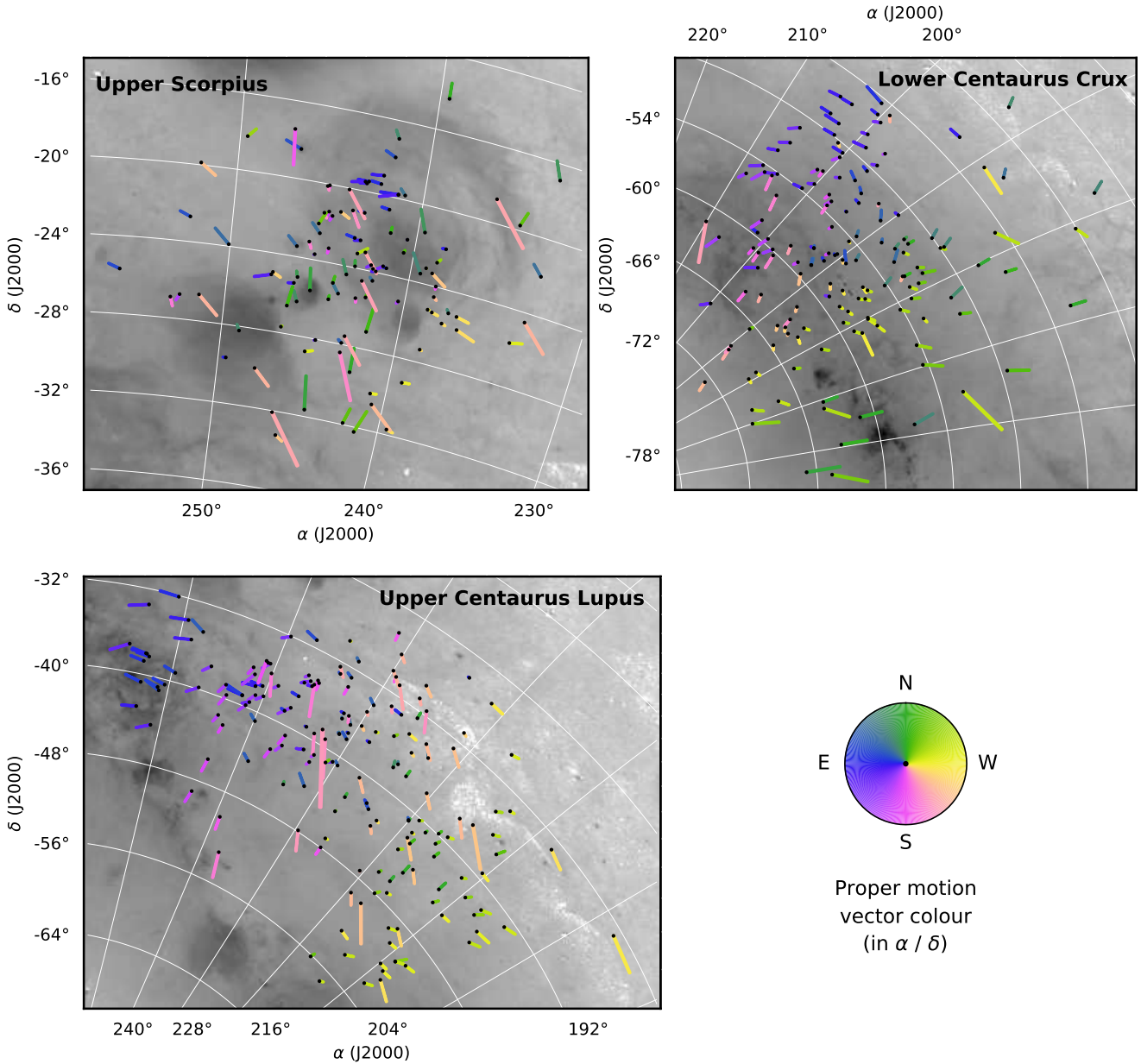}
\caption{Proper motion vector maps for the three subgroups of Sco-Cen projected onto inverted H$\alpha$ images from \citet{fink03}. The proper motions have been corrected for radial streaming motion according to the equations in \citet{brow97} and using bulk (median) RVs of -6.2, 2.8 and 13.0 km/s for the three subgroups. Points show the current positions of stars and vectors show the proper motions over 0.5~Myr with the bulk motion of each subgroup subtracted to show the motion in the reference frame of each subgroup. The vectors have been colour-coded based on their direction of motion (see colour wheel in lower-right corner) to highlight kinematic substructure. Note that because the kinematic reference frame shown for each subgroup is different, the colours of the proper motions in each subgroup do not represent the same absolute proper motion as stars in other subgroups.}
\label{corrected_PMs}
\end{center}
\end{figure*}

\subsection{Corrected proper motion vector maps}
\label{s-correctedPMs}

If a group of stars has a non-zero radial motion this will cause an apparent expansion or contraction of the group in their proper motions as their spatial extent on the sky changes \citep[referred to as {\it virtual expansion / contraction},][]{brow97}. If the bulk RV of the stars is known one can perform a correction for this virtual expansion / contraction, effectively calculating the contribution to the proper motions due to the radial motion of the association and then subtracting these from the observed proper motions to provide corrected proper motions. To do this we use the equations in Appendix A1 of \citet{brow97}, using the median RV of stars in each subgroup (to minimise the impact of binarity).

The corrected proper motion vector maps for the three subgroups are shown in Figure~\ref{corrected_PMs}. None of the three subgroups show evidence for a coherent expansion pattern. UCL does appear to show some coherent patterns consistent with expansion, with stars in the north-east and south-west appearing to move away from the centre of the subgroup, though stars elsewhere in the group do not show this behaviour. We will argue, in Section~\ref{s-kinematicsubstructure}, that these patterns are due to the kinematic substructure within the subgroups.

To quantify the expansion of each subgroup we follow the method of \citet{wrig16} and calculate, for each star, the amount of kinetic energy in the radial and transverse directions (using the subgroup centres calculated in Section~\ref{s-spatialmodel}), dividing the former between expansion and contraction, and then sum these for each subgroup to provide global energy fractions (for simplicity we assume all stars have equal masses). If a subgroup was expanding radially from a single point we would expect the majority of kinetic energy to be in the radial direction and for the stars to be moving outwards.

We find that the kinetic energy is divided almost equally between the radial and transverse directions, with 50\% (US), 66\% (UCL), and 45\% (LCC) of energy in the radial direction. Only UCL exhibits a preference for kinetic energy in the radial direction. In the radial direction there is a preference for the motions of stars to be directed away from the centres of each subgroup, with 59\% (US), 67\% (UCL) and 90\% (LCC) of the kinetic energy in the direction of expansion. This suggests that, while there isn't evidence for coherent expansion patterns in the correct proper motions of stars in each subgroup, they do appear to be expanding.

\section{Kinematic substructure in the association}
\label{s-kinematicsubstructure}

\begin{figure*}
\begin{center}
\includegraphics[width=510pt]{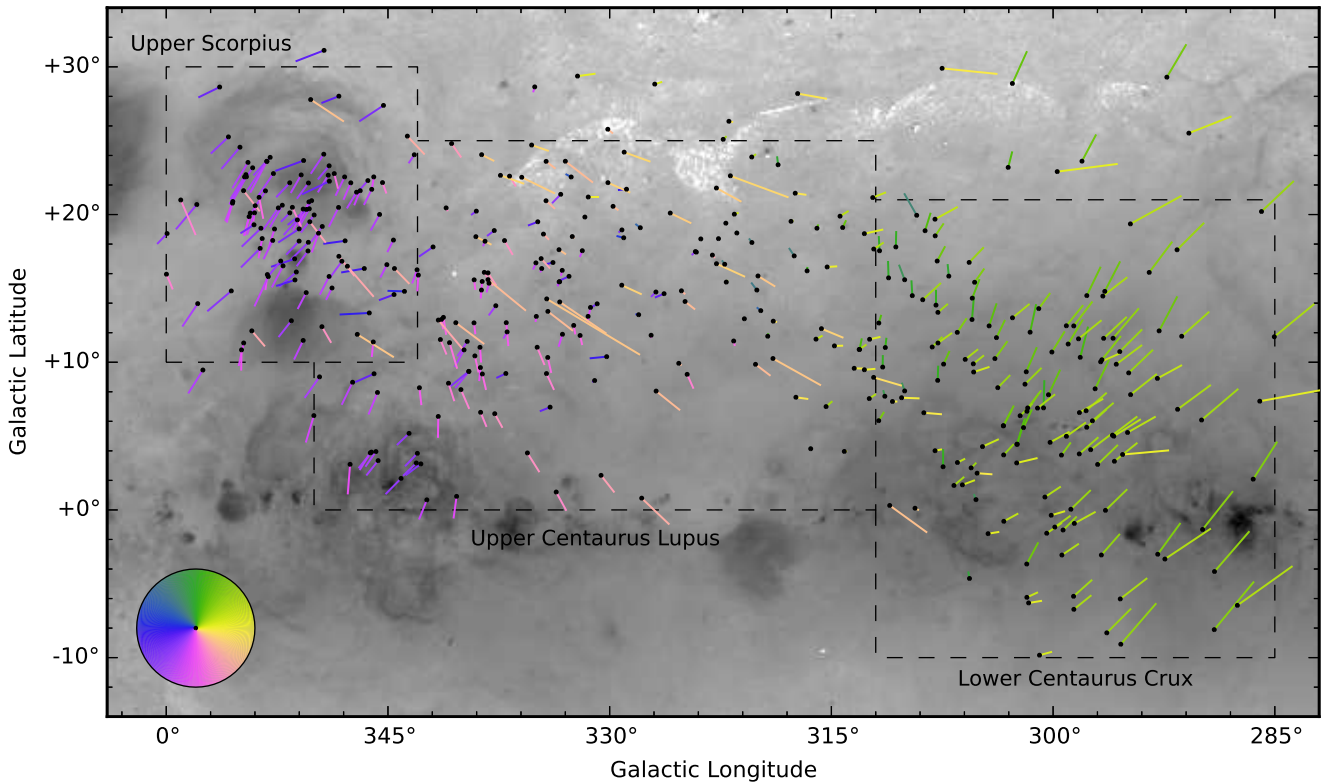}
\caption{Proper motion vector map for the entire Sco-Cen association projected onto an inverted H$\alpha$ image from \citet{fink03}. The proper motions were corrected for radial streaming motion according to the equations in \citet{brow97} and using bulk (median) RVs of -6.2, 2.8 and 13.0 km/s for the three subgroups. Points show the current positions of stars and vectors show the proper motions over 0.5~Myr with the bulk (median) proper motion of the entire association subtracted to show the relative motion. The vectors have been colour based on their direction of motion (see colour wheel in lower-left) to highlight kinematic substructure.}
\label{map_substructure}
\end{center}
\end{figure*}

Kinematic substructure is the observed tendency for stars in the same area of a star forming region or association to have motions that are more similar to their neighbours than to other stars. It may manifest as sizeable substructures within the association \citep[e.g., the Gamma Velorum cluster shows two distinct kinematic components,][]{jeff14} or as smaller, more numerous substructures \citep[as observed in Cygnus OB2,][]{wrig16}. In the case of OB associations, kinematic substructure is believed to be a remnant of the primordial phase-space structure that was present when the association formed \citep[e.g.,][]{lars81,wrig16}. In this Section we explore the evidence for kinematic substructure in Sco-Cen.

Figure~\ref{map_substructure} shows a corrected proper motion vector map for Sco-Cen with the vectors colour-coded by the position angle of their motion on the celestial sphere \citep[as per][]{wrig16}. In this figure the three subgroups stand out relatively clearly due to their different kinematics, with some overlap between the groups that suggests their membership could benefit from some adjustments.

In addition to the three subgroups there is also a group of stars visible in this diagram with above average velocities moving in the direction of decreasing latitude and longitude (they appear a creamy-orange colour in this diagram). These stars have different proper motions from the stars they are projected against because, despite having similar kinematics, they are in the foreground at distances of 50--100~pc. Four of these stars are significantly in the foreground ($d < 70$~pc) and so are unlikely to be members of Sco-Cen; HIP 68413, HIP 69995, HIP 70931 and HIP 76063. We can find no evidence in the literature that these stars are particularly young \citep[e.g.,][]{davi15} and therefore suggest that these are field stars that should be removed from the membership of Sco-Cen.

Figure~\ref{corrected_PMs} shows corrected proper motion vector maps for the three subgroups of Sco-Cen. From this it is evident that the proper motions of stars in each subgroup are not random but are correlated with position. Each subgroup shows evidence for multiple groups of stars each in their own area of the sky and each with stars that are all moving in a similar direction. UCL and LCC are particularly noteworthy as they each exhibit a small number of groups ($\sim$4--5) each with 10--30 members and in a distinct part of the association. In UCL and LCC there are hints of expansion in the $\alpha$ axis in that stars in the east of the subgroups are moving eastwards and stars in the west are moving westwards. Since this axis is close to the Galactic $Y$ axis (in UCL and LCC) this is probably the same feature we observed in Section~\ref{s-3Dexpansion}. In US the kinematic substructures are not as distinct and the subgroup appears to be made up of a larger number of smaller substructures, similar to the kinematic substructure observed in Cygnus~OB2.

\subsection{Quantifying kinematic substructure}

\subsubsection{Method}

In an attempt to quantify this substructure we follow \citet{wrig16} and use Moran's spatial correlation index I \citep{mora50}, which quantifies the overall degree of similarity between spatially close regions with respect to a numeric variable (in this case the proper motion in either dimension). The expectation value for no spatial correlation is $I_0 = -1 / (n-1)$, with $I > I_0$ indicating positive spatial correlation and $I < I_0$ indicating negative spatial correlation. We use this instead of Geary's C index \citep{gear54}, since Moran's I statistic is a global measure that can identify and quantify large-scale spatial correlation (as appears to be evident in Figure~\ref{corrected_PMs}), while Geary's C is a local measure of spatial correlation. For brevity we don't provide quantification of the index, but instead refer the reader to these papers and \citet{wrig16}.

We calculate the index for the proper motions in both directions for all three subgroups of Sco-Cen, using all the stars in our sample. We did this in proper motion space rather than in cartesian $UVW$ space as this would have required the use of individual RVs that are distorted by binary motions leading to an artificial `smearing' of the phase-space structure in this direction. We take into account the effect that the projection of motions onto a curved surface have on the measured kinematic substructure by creating synthetic populations of each subgroup (consisting of stars at the observed positions and distances of each star but with proper motions calculated from a uniform $UVW$ space motion with an added velocity dispersion equal to those measured in Section~\ref{s-velocitymodel}). We then calculated Moran's $I$ index for the simulated dataset and repeated this 10,000 times for each subgroup to determine the typical level of kinematic substructure that would be measured solely due to this.

\subsubsection{Results}

For US we calculate values of $I_\alpha = 0.135 \pm 0.048$ and $I_\delta = -0.013 \pm 0.048$, indicating large and significant (3.0$\sigma$) positive spatial correlation (nearby regions tend to exhibit similar velocities) for $\mu_\alpha$, and a very weak measure of negative spatial correlation (nearby regions exhibit dissimilar velocities) for $\mu_\delta$ that is consistent with no spatial correlation. The large difference between the two dimensions is probably because the compact kinematic substructures evident in US (such as the groups at $\alpha$, $\delta$ = 243, -20 or 237, -25) have most of their motion in $\alpha$ leading to significant spatial correlation in $\alpha$ but not $\delta$. Our Monte Carlo simulations indicate that the projection of a bulk motion into proper motions would give typical values of $I_\alpha = 0.002$ (0.24$\sigma$) and $I_\delta = 0.006$ (0.31$\sigma$), which at least for the $\alpha$ dimension is significantly smaller than the measured value.

For UCL we calculate values of $I_\alpha = 0.000 \pm 0.022$ and $I_\delta = -0.034 \pm 0.022$, indicating weak spatial correlation in both dimensions that is  consistent with no spatial correlation in $\mu_\alpha$ or has a low significance (1.3$\sigma$) of negative spatial correlation in $\mu_\delta$. The latter is likely due to the presence of an extended group of rapidly moving stars in pink in Figure~\ref{corrected_PMs} that are moving in a southerly direction. The presence of these stars in between stars moving in different directions will cause negative spatial correlation, particularly in $\mu_\delta$. Our Monte Carlo simulations predict that with no kinematic substructure we should measure substructure with significances of 0.21 and 0.11$\sigma$, which in $\alpha$ is consistent with the level measured.

For LCC we calculate $I_\alpha = 0.024 \pm 0.019$ and $I_\delta = 0.029 \pm 0.019$, indicating positive spatial correlation in both dimensions at 1.6 and 1.9$\sigma$, respectively. Our Monte Carlo simulations predict that with no kinematic substructure we should measure substructure with significances of 0.40 and 0.31$\sigma$, which slightly reduces the impact of the kinematic substructure we have measured.

Taken together these results suggest that while there is kinematic substructure in each of the three subgroups of Sco-Cen, the level of kinematic substructure varies between subgroups and is lower than that measured in Cyg~OB2 \citep{wrig16}.

\section{Discussion}

\subsection{The virial state of Sco-Cen's subgroups}
\label{r-virialstate}

To determine the virial state of each subgroup of Sco-Cen we use the virial equation, which in its 3D form is given by

\begin{equation}
\sigma_{3D}^2 = \frac{G M_{vir}}{2 r_{vir}}
\end{equation}

\noindent where $\sigma_{3D}$ is the 3D velocity dispersion, $G$ is the gravitational constant, $M_{vir}$ is the virial mass, and $r_{vir}$ is the virial radius. We substitute the parameter $\eta = 6 r_{vir} / r_{eff}$ \citep[e.g.,][]{port10}, where $r_{eff}$ is the effective (or half-light) radius, giving

\begin{equation}
M_{vir} = \eta \, \frac{\sigma_{3D}^2 \, r_{eff}}{3G}
\end{equation}

\noindent where the parameters $\eta$ and $r_{eff}$ can be calculated from the \citet{elso87} parameters $\gamma$ and $a$ (see Section~\ref{s-spatialmodel}). For the large $\gamma$ values calculated for the subgroups we find $\eta$ values of 9.1 (US), 9.0 (UCL), and 9.1 (LCC)\footnote{The dependence of $\eta$ on $\gamma$ is very weak at large $\gamma$ values and therefore the uncertainties on $\eta$ are almost zero.}, while the effective radii, $r_{eff}$ are $11.3^{+0.5}_{-0.3}$ (US), $20.0^{+0.7}_{-0.7}$ (UCL), and $15.5^{+0.9}_{-0.8}$~pc (LCC).

Combining the effective radii, 3D velocity dispersions (from Table~\ref{structure}), and $\eta$ values, we calculate virial masses of $8.2^{+1.5}_{-1.2} \times 10^4$ (US), $8.3^{+1.8}_{-1.5} \times 10^4$ (UCL), and $5.1^{+2.9}_{-1.3} \times 10^4$ M$_\odot$ (LCC). These values can be compared to the total mass of stars in each subgroup to estimate the virial state of each group. \citet{prei08} estimate US to have a total mass of $\sim$2000~M$_\odot$, while \citet{mama02} estimate UCL and LCC to consist of approximately 2200 and 1200 stars, respectively, which equates to total stellar masses of about 1250 and 700 M$_\odot$ \citep[using the same initial mass function and binary fraction as][]{prei08}. All of these estimates are significantly smaller than the calculated virial masses, by at least an order of magnitude, implying that the three subgroups are in a supervirial state, as expected given their low stellar density. We would therefore expect the subgroups to be in a state of expansion.

It is interesting to note that the virial mass estimated for US is very close to the mass of atomic H~{\sc i} in the shell surrounding the subgroup, which \citet{dege92} estimate to have a mass of $8 \times 10^4$~M$_\odot$. They argue that the shell consists of swept-up material from the primordial GMC from which US formed, which suggests that the subgroup could have been born close to virial equilibrium within the GMC (if the stars would have felt the gravitational potential of the entire GMC). The masses of the shells surrounding UCL and LCC, $3 \times 10^5$ and $1 \times 10^5$~M$_\odot$ respectively, are more massive than their virial masses, implying that the subgroups might have initially been in a subvirial state.

\subsection{Expansion of the Sco-Cen subgroups}

Ever since \citet{amba47} we have known that OB associations are unbound and will therefore expand. This has lead many authors to postulate that they were more compact in the past and have been expanding for a while \citep[e.g.,][]{blaa64,blaa91,brow99}. These ideas lead to suggestions that OB associations may be the expanded remnants of compact star clusters \citep{lada03}, disrupted by processes such as residual gas expulsion \citep[e.g.,][]{hill80,lada84}.

In Section~\ref{s-expansion} we used five different methods to determine whether the subgroups of Sco-Cen are expanding, and if so whether they are expanding from a previously compact configuration such as a dense star cluster. Many of these methods assume that the cluster would expand radially (such as the expansion models explored in Sections \ref{s-linearexpansion} and \ref{s-3Dexpansion}), while others better account for the Galactic potential (Section~\ref{s-traceback}) or provide an overall view of the stellar motions (Section~\ref{s-correctedPMs}). Some of these methods use individual RVs (where available), while others use the bulk RV (and are thus less affected by binary motions).

None of the methods provide evidence that the subgroups are expanding from a more compact configuration. The linear expansion model considered in Section~\ref{s-linearexpansion} finds negative slopes that imply contraction rather than expansion, though all three are consistent with no expansion or contraction to $\sim$1$\sigma$. The 3D expansion model (Section~\ref{s-3Dexpansion}) again shows slopes consistent with no expansion or contraction, except along the $Y$ axis where all three subgroups exhibit evidence for expansion to a confidence of 2--3$\sigma$. Interestingly, \citet{mama14} found for the $\beta$ Pictoris moving group that the most significant evidence for expansion also came in the $Y$ direction. This could be caused by galactic shear since the surface densities of OB associations like Sco-Cen are low enough for shear to be effective, though the timescales involved are probably too short \citep[][estimate a shear timescale of $\sim$70~Myr in the vicinity of the Sun]{dobb13}. Alternatively this shear pattern may have been imprinted on the molecular gas in the primordial GMC and then inherited by the OB association that formed. Since the OB association does not appear to be dynamically evolved such a kinematic pattern could have survived the early evolution of the system.

By tracing back the stellar motions (Section~\ref{s-traceback}) there is evidence that the subgroups were actually larger in the past than they are today. This result is supported by the corrected proper motion vector maps in Figure~\ref{corrected_PMs} that show no preference for motions radially away from the subgroup centres, though amongst the radial motions there is a preference for expanding motions over contracting motions. Similar results have been found for other associations and moving groups, such as the $\beta$ Pictoris moving group \citep{mama14} and Tuc-Hor \citep{maka07}, both of which do not appear to have been significantly smaller in the past.

In summary, while the three subgroups are all gravitationally unbound and will therefore expand in the future, they all lack evidence for having been in a more compact configuration in the past. This goes against the view that associations expand as they age from an almost universal compact configuration \citep[e.g.,][]{pfal09}. This is particularly clear in the case of the three subgroups of Sco-Cen, which \citet{pfal13} suggest had an original size of 1--3~pc, approximately an order of magnitude smaller than their current size. The kinematics of stars in the subgroups shows that this is definitely not the case. Furthermore there is no evidence for a coherent radial expansion pattern amongst the members of each subgroup and the kinematic substructure evident in Figure~\ref{corrected_PMs} suggest a more complex expansion pattern consisting of multiple expanding substructures within each subgroup.

\subsection{Kinematic substructure and the formation of Sco-Cen}

Molecular clouds are known to be highly substructured, both spatially \citep{falg91} and kinematically \citep{haca13}, with this hierarchical or fractal structure passed on to the formed stars \citep{elme01} and often evident in their initial spatial distribution \citep{gute08}. Dynamical interactions between groups of stars can erase this substructure \citep[e.g.,][]{scal02,park14}, a process that acts on the dynamical timescale of a region. If the dynamical timescale is longer than the age of the region then the primordial substructure can be preserved \citep[e.g.,][]{good04}, though dynamical timescales can change as a region expands or collapses (so an instantaneous measure of the dynamical timescale isn't always so revealing). The amount of substructure present in a region can constrain the past dynamical evolution of a group of stars, since it requires that a region cannot have had periods with a short dynamical time otherwise dynamical mixing would have erased its substructure. Compact star clusters have significantly shorter dynamical timescales than extended OB associations, so the presence of physical or dynamical substructure in a region means it can never have been in a highly compact phase in the past \citep[e.g.,][]{park14}.

The three subgroups of Sco-Cen each display evidence for kinematic substructure, as has been observed in other OB associations \citep{wrig14b} and across large star forming complexes \citep[e.g.,][]{fure06,fure08,tobi15}. US exhibits kinematic substructure similar to that seen in Cyg~OB2, with many small subgroups containing 5--10 stars from our sample, and some stars that don't appear to be part of any visible substructures, while UCL and LCC exhibit much larger substructures with 10s of stars in each group, but only 2--4 clear groups in each subgroup. These substructures may be responsible for the complex star formation history revealed by \citet{peca16}, with the observed age spreads within each subgroup due to the combination of individual substructures within it, each of which may not have a significant age spread. We therefore agree with the conclusions of those authors that the current division of Sco-Cen into three subgroups is too simplistic and at least UCL and LCC should be broken down into smaller subgroups based on their ages or kinematics.

A number of different formation scenarios have been considered for Sco-Cen, ranging from the sequential star formation process originally proposed by \citet{blaa64} and extended by \citet{prei99}, the impact of a high-velocity cloud on the disk \citep{lepi94}, to its formation in a large ring-like structure containing other young groups of stars known as the Gould Belt \citep{lind73}. The high-velocity cloud scenario and the Gould Belt model both predict stellar motions that are opposite to those observed and so have been ruled out \citep[e.g.,][]{sart03,prei08}. The sequential star formation model suggests that star formation began in UCL and propagated to LCC, US, and then the $\rho$~Ophiuchus star forming region through triggering by supernova shockwaves. The presence of kinematic substructure and age spreads within the subgroups that suggests they would be better subdivided into smaller groups with a more complex star formation history complicates this model, but doesn't necessarily disprove it.

\section{Conclusions}

Our kinematic study of the Sco-Cen OB association using {\it Gaia} DR1 parallaxes and proper motions has lead to the following findings.

\begin{enumerate}
\item We use Bayesian inference and forward modelling to calculate 3D velocity dispersions for the three subgroups. All three have non-isotropic velocity dispersions, suggesting they are not dynamically relaxed and likely have not been in the past. The 3D velocity dispersions are $3.20^{+0.22}_{-0.20}$ (US), $2.45^{+0.20}_{-0.20}$ (UCL), and $2.15^{+0.47}_{-0.24}$~km/s (LCC). These imply virial masses that are over an order of magnitude larger than the known stellar mass, confirming the subgroups are gravitationally unbound.
\item We have explored multiple methods for testing if and how the subgroups are expanding, including the \citet{blaa64b} linear expansion model, 3D linear expansion, a comparison of expanding and non-expanding convergence points, tracing back the individual stellar motions to identify the smallest size of each subgroup in the past, and studying corrected proper motion vector maps. The kinematic data are inconsistent with the subgroups being the expanded remnants of individual star clusters, with no coherent expansion pattern evident and no evidence that the subgroups had a more compact configuration in the past.
\item The subgroups all show evidence for kinematic substructure, which we quantify using a spatial correlation test. US appears to consist of multiple small substructures, similar to that observed in Cyg~OB2 \citep{wrig16}, thought not as strong, while UCL and LCC appear to be made up of a smaller number of larger substructures with a lower significance. The presence of these substructures places constraints on the past structure and dynamical evolution of the subgroups.
\end{enumerate}

To conclude, the three subgroups of Sco-Cen are not the expanded remnants of individual star clusters and instead are likely to have formed with considerable physical and kinematic sub- structure, such as from numerous smaller clusters. Much of this kinematic substructure remains today, most likely because the subgroups have not undergone a densely clustered phase during which the substructure would have been erased. US retains considerable substructure that may be very similar to its primordial structure, while UCL and LCC appear to consist of a few larger substructures that may indicate either that they were born as such groups or that these substructures have undergone some mergers already. Combined with the recent evidence from \citet{peca16} showing that Sco-Cen has considerable substructure in its age distribution (which is likely to be closely related to its kinematic substructure) our results suggest that the subgroups did not form as individual bursts of star formation but are instead composed of multiple smaller subgroups, each of which probably formed in a different place and at a different time.

Combined with recent studies \citep[e.g.,][]{maka07,mama14,wrig16} our results suggest that the classical view of OB associations (and moving groups) being the expanded remnants of star clusters is incorrect. Instead these groups appear to have formed with more extended and substructured spatial and kinematic distributions. This implies that the assumption that clusters and associations expand uniformly as they age is not true \citep{pfal09} and that star cluster disruption mechanisms, such as residual gas expulsion \citep{krou01}, are not necessary for dispersing the majority of young stars into the Galactic field. This has significant implications for the formation of the Galactic field and the birth environment of stars and proto-planetary disks.

\section{Acknowledgments}

NJW acknowledges an STFC Ernest Rutherford Fellowship (grant number ST/M005569/1). Part of this work was completed at the Jet Propulsion Laboratory, California Institute of Technology, under a contract with the National Aeronautics and Space Administration. The authors would like to thank Mark Pecaut and Rob Jeffries for discussions and comments on this paper, as well as the anonymous referee for comments and suggestions that improved the quality of this paper. This work has made use of data from the ESA space mission Gaia (http://www.cosmos.esa.int/gaia), processed by the Gaia Data Processing and Analysis Consortium (DPAC, http://www.cosmos.esa.int/web/gaia/dpac/consortium). Funding for DPAC has been provided by national institutions, in particular the institutions participating in the Gaia Multilateral Agreement. This research made use of the Simbad and Vizier catalogue access tools (provided by CDS, Strasbourg, France), Astropy \citep{astr13} and TOPCAT \citep{tayl05}. 

\bibliographystyle{mn2e}
\bibliography{/Users/nwright/Documents/Work/tex_papers/bibliography.bib}

\begin{thebibliography}{103}
\expandafter\ifx\csname natexlab\endcsname\relax\def\natexlab#1{#1}\fi

\bibitem[{{Adams} {et~al}\mbox{.}(2006){Adams}, {Proszkow}, {Fatuzzo}, \&
  {Myers}}]{adam06}
{Adams} F.~C., {Proszkow} E.~M., {Fatuzzo} M., {Myers} P.~C., 2006, \apj, 641,
  504

\bibitem[{{Ambartsumian}(1947)}]{amba47}
{Ambartsumian} V.~A., 1947, {Stellar Evolution and Astrophysics}. Armenian
  Acad. of Sci.

\bibitem[{{Anderson} \& {Francis}(2012)}]{ande12}
{Anderson} E., {Francis} C., 2012, Astronomy Letters, 38, 331

\bibitem[{{Arenou} {et~al}\mbox{.}(2017){Arenou}, {Luri}, {Babusiaux},
  {Fabricius}, {Helmi}, {Robin}, {Vallenari}, {Blanco-Cuaresma},
  {Cantat-Gaudin}, {Findeisen}, {Reyl{\'e}}, {Ruiz-Dern}, {Sordo}, {Turon},
  {Walton}, {Shih}, {Antiche}, {Barache}, {Barros}, {Breddels}, {Carrasco},
  {Costigan}, {Diakit{\'e}}, {Eyer}, {Figueras}, {Galluccio}, {Heu}, {Jordi},
  {Krone-Martins}, {Lallement}, {Lambert}, {Leclerc}, {Marrese}, {Moitinho},
  {Mor}, {Romero-G{\'o}mez}, {Sartoretti}, {Soria}, {Soubiran}, {Souchay},
  {Veljanoski}, {Ziaeepour}, {Giuffrida}, {Pancino}, \& {Bragaglia}}]{aren17}
{Arenou} F. {et~al.}, 2017, \aap, 599, A50

\bibitem[{{Astraatmadja} \& {Bailer-Jones}(2016)}]{astr16}
{Astraatmadja} T.~L., {Bailer-Jones} C.~A.~L., 2016, \apj, 832, 137

\bibitem[{{Astropy Collaboration} {et~al}\mbox{.}(2013){Astropy Collaboration},
  {Robitaille}, {Tollerud}, {Greenfield}, {Droettboom}, {Bray}, {Aldcroft},
  {Davis}, {Ginsburg}, {Price-Whelan}, {Kerzendorf}, {Conley}, {Crighton},
  {Barbary}, {Muna}, {Ferguson}, {Grollier}, {Parikh}, {Nair}, {Unther},
  {Deil}, {Woillez}, {Conseil}, {Kramer}, {Turner}, {Singer}, {Fox}, {Weaver},
  {Zabalza}, {Edwards}, {Azalee Bostroem}, {Burke}, {Casey}, {Crawford},
  {Dencheva}, {Ely}, {Jenness}, {Labrie}, {Lim}, {Pierfederici}, {Pontzen},
  {Ptak}, {Refsdal}, {Servillat}, \& {Streicher}}]{astr13}
{Astropy Collaboration} {et~al.}, 2013, \aap, 558, A33

\bibitem[{{Bailer-Jones}(2015)}]{bail15}
{Bailer-Jones} C.~A.~L., 2015, \pasp, 127, 994

\bibitem[{{Baumgardt} \& {Kroupa}(2007)}]{baum07}
{Baumgardt} H., {Kroupa} P., 2007, \mnras, 380, 1589

\bibitem[{{Blaauw}(1946)}]{blaa46}
{Blaauw} A., 1946, Publications of the Kapteyn Astronomical Laboratory
  Groningen, 52, 1

\bibitem[{{Blaauw}(1964{\natexlab{a}})}]{blaa64}
{Blaauw} A., 1964{\natexlab{a}}, \araa, 2, 213

\bibitem[{{Blaauw}(1964{\natexlab{b}})}]{blaa64b}
{Blaauw} A., 1964{\natexlab{b}}, in IAU Symposium, Vol.~20, The Galaxy and the
  Magellanic Clouds, {Kerr} F.~J., ed., p.~50

\bibitem[{{Blaauw}(1991)}]{blaa91}
{Blaauw} A., 1991, in NATO ASIC Proc. 342: The Physics of Star Formation and
  Early Stellar Evolution, {Lada} C.~J., {Kylafis} N.~D., eds., pp. 125--+

\bibitem[{{Breitschwerdt} {et~al}\mbox{.}(2016){Breitschwerdt}, {Feige},
  {Schulreich}, {Avillez}, {Dettbarn}, \& {Fuchs}}]{brei16}
{Breitschwerdt} D., {Feige} J., {Schulreich} M.~M., {Avillez} M.~A.~D.,
  {Dettbarn} C., {Fuchs} B., 2016, \nat, 532, 73

\bibitem[{{Bressert} {et~al}\mbox{.}(2010){Bressert}, {Bastian}, {Gutermuth},
  {Megeath}, {Allen}, {Evans}, {Rebull}, {Hatchell}, {Johnstone}, {Bourke},
  {Cieza}, {Harvey}, {Merin}, {Ray}, \& {Tothill}}]{bres10}
{Bressert} E. {et~al.}, 2010, \mnras, 409, L54

\bibitem[{{Brown} {et~al}\mbox{.}(1999){Brown}, {Blaauw}, {Hoogerwerf}, {de
  Bruijne}, \& {de Zeeuw}}]{brow99}
{Brown} A.~G.~A., {Blaauw} A., {Hoogerwerf} R., {de Bruijne} J.~H.~J., {de
  Zeeuw} P.~T., 1999, in NATO Advanced Science Institutes (ASI) Series C,
  {Lada} C.~J., {Kylafis} N.~D., eds., Vol. 540, p. 411

\bibitem[{{Brown} {et~al}\mbox{.}(1997){Brown}, {Dekker}, \& {de
  Zeeuw}}]{brow97}
{Brown} A.~G.~A., {Dekker} G., {de Zeeuw} P.~T., 1997, \mnras, 285, 479

\bibitem[{{Carpenter} {et~al}\mbox{.}(2006){Carpenter}, {Mamajek},
  {Hillenbrand}, \& {Meyer}}]{carp06}
{Carpenter} J.~M., {Mamajek} E.~E., {Hillenbrand} L.~A., {Meyer} M.~R., 2006,
  \apjl, 651, L49

\bibitem[{{Chen} {et~al}\mbox{.}(2011){Chen}, {Mamajek}, {Bitner}, {Pecaut},
  {Su}, \& {Weinberger}}]{chen11}
{Chen} C.~H., {Mamajek} E.~E., {Bitner} M.~A., {Pecaut} M., {Su} K.~Y.~L.,
  {Weinberger} A.~J., 2011, \apj, 738, 122

\bibitem[{{Cottaar} {et~al}\mbox{.}(2012){Cottaar}, {Meyer}, {Andersen}, \&
  {Espinoza}}]{cott12}
{Cottaar} M., {Meyer} M.~R., {Andersen} M., {Espinoza} P., 2012, \aap, 539, A5

\bibitem[{{Dahm} {et~al}\mbox{.}(2012){Dahm}, {Slesnick}, \& {White}}]{dahm12}
{Dahm} S.~E., {Slesnick} C.~L., {White} R.~J., 2012, \apj, 745, 56

\bibitem[{{David} \& {Hillenbrand}(2015)}]{davi15}
{David} T.~J., {Hillenbrand} L.~A., 2015, \apj, 804, 146

\bibitem[{{de Bruijne}(1999{\natexlab{a}})}]{debr99b}
{de Bruijne} J.~H.~J., 1999{\natexlab{a}}, \mnras, 306, 381

\bibitem[{{de Bruijne}(1999{\natexlab{b}})}]{debr99}
{de Bruijne} J.~H.~J., 1999{\natexlab{b}}, \mnras, 310, 585

\bibitem[{{de Geus}(1992)}]{dege92}
{de Geus} E.~J., 1992, \aap, 262, 258

\bibitem[{{de Geus} {et~al}\mbox{.}(1989){de Geus}, {de Zeeuw}, \&
  {Lub}}]{dege89}
{de Geus} E.~J., {de Zeeuw} P.~T., {Lub} J., 1989, \aap, 216, 44

\bibitem[{{de Zeeuw} {et~al}\mbox{.}(1999){de Zeeuw}, {Hoogerwerf}, {de
  Bruijne}, {Brown}, \& {Blaauw}}]{deze99}
{de Zeeuw} P.~T., {Hoogerwerf} R., {de Bruijne} J.~H.~J., {Brown} A.~G.~A.,
  {Blaauw} A., 1999, \aj, 117, 354

\bibitem[{{Dobbs} \& {Pringle}(2013)}]{dobb13}
{Dobbs} C.~L., {Pringle} J.~E., 2013, \mnras, 432, 653

\bibitem[{{Duch{\^e}ne} \& {Kraus}(2013)}]{duch13}
{Duch{\^e}ne} G., {Kraus} A., 2013, \araa, 51, 269

\bibitem[{{Elmegreen} {et~al}\mbox{.}(2000){Elmegreen}, {Efremov}, {Pudritz},
  \& {Zinnecker}}]{elme00}
{Elmegreen} B.~G., {Efremov} Y., {Pudritz} R.~E., {Zinnecker} H., 2000,
  Protostars and Planets IV, 179

\bibitem[{{Elmegreen} \& {Elmegreen}(2001)}]{elme01}
{Elmegreen} B.~G., {Elmegreen} D.~M., 2001, \aj, 121, 1507

\bibitem[{{Elson} {et~al}\mbox{.}(1987){Elson}, {Fall}, \& {Freeman}}]{elso87}
{Elson} R.~A.~W., {Fall} S.~M., {Freeman} K.~C., 1987, \apj, 323, 54

\bibitem[{{ESA}(1997)}]{esa97}
{ESA}, ed., 1997, ESA Special Publication, Vol. 1200, {The HIPPARCOS and TYCHO
  catalogues. Astrometric and photometric star catalogues derived from the ESA
  HIPPARCOS Space Astrometry Mission}. {ESA}

\bibitem[{{Falgarone} {et~al}\mbox{.}(1991){Falgarone}, {Phillips}, \&
  {Walker}}]{falg91}
{Falgarone} E., {Phillips} T.~G., {Walker} C.~K., 1991, \apj, 378, 186

\bibitem[{{Feast} \& {Whitelock}(1997)}]{feas97}
{Feast} M., {Whitelock} P., 1997, \mnras, 291, 683

\bibitem[{{F{\H u}r{\'e}sz} {et~al}\mbox{.}(2008){F{\H u}r{\'e}sz}, {Hartmann},
  {Megeath}, {Szentgyorgyi}, \& {Hamden}}]{fure08}
{F{\H u}r{\'e}sz} G., {Hartmann} L.~W., {Megeath} S.~T., {Szentgyorgyi} A.~H.,
  {Hamden} E.~T., 2008, \apj, 676, 1109

\bibitem[{{F{\H u}r{\'e}sz} {et~al}\mbox{.}(2006){F{\H u}r{\'e}sz}, {Hartmann},
  {Szentgyorgyi}, {Ridge}, {Rebull}, {Stauffer}, {Latham}, {Conroy},
  {Fabricant}, \& {Roll}}]{fure06}
{F{\H u}r{\'e}sz} G. {et~al.}, 2006, \apj, 648, 1090

\bibitem[{{Finkbeiner}(2003)}]{fink03}
{Finkbeiner} D.~P., 2003, \apjs, 146, 407

\bibitem[{{Foreman-Mackey} {et~al}\mbox{.}(2013){Foreman-Mackey}, {Hogg},
  {Lang}, \& {Goodman}}]{fore13}
{Foreman-Mackey} D., {Hogg} D.~W., {Lang} D., {Goodman} J., 2013, \pasp, 125,
  306

\bibitem[{{Fuchs} {et~al}\mbox{.}(2006){Fuchs}, {Breitschwerdt}, {de Avillez},
  {Dettbarn}, \& {Flynn}}]{fuch06}
{Fuchs} B., {Breitschwerdt} D., {de Avillez} M.~A., {Dettbarn} C., {Flynn} C.,
  2006, \mnras, 373, 993

\bibitem[{{Gaia Collaboration} {et~al}\mbox{.}(2016{\natexlab{a}}){Gaia
  Collaboration}, {Brown}, {Vallenari}, {Prusti}, {de Bruijne}, {Mignard},
  {Drimmel}, {Babusiaux}, {Bailer-Jones}, {Bastian}, \& et~al.}]{brow16}
{Gaia Collaboration} {et~al.}, 2016{\natexlab{a}}, \aap, 595, A2

\bibitem[{{Gaia Collaboration} {et~al}\mbox{.}(2016{\natexlab{b}}){Gaia
  Collaboration}, {Prusti}, {de Bruijne}, {Brown}, {Vallenari}, {Babusiaux},
  {Bailer-Jones}, {Bastian}, {Biermann}, {Evans}, \& et~al.}]{prus16}
{Gaia Collaboration} {et~al.}, 2016{\natexlab{b}}, \aap, 595, A1

\bibitem[{{Geary}(1954)}]{gear54}
{Geary} R.~C., 1954, The Incorporated Statistician, 5, 115

\bibitem[{{Gontcharov}(2006)}]{gont06}
{Gontcharov} G.~A., 2006, Astronomy Letters, 32, 759

\bibitem[{{Goodwin} \& {Whitworth}(2004)}]{good04}
{Goodwin} S.~P., {Whitworth} A.~P., 2004, \aap, 413, 929

\bibitem[{{Guarcello} {et~al}\mbox{.}(2016){Guarcello}, {Drake}, {Wright},
  {Albacete-Colombo}, {Clarke}, {Ercolano}, {Flaccomio}, {Kashyap}, {Micela},
  {Naylor}, {Schneider}, {Sciortino}, \& {Vink}}]{guar16}
{Guarcello} M.~G. {et~al.}, 2016, ArXiv e-prints

\bibitem[{{Gutermuth} {et~al}\mbox{.}(2008){Gutermuth}, {Myers}, {Megeath},
  {Allen}, {Pipher}, {Muzerolle}, {Porras}, {Winston}, \& {Fazio}}]{gute08}
{Gutermuth} R.~A. {et~al.}, 2008, \apj, 674, 336

\bibitem[{{Hacar} {et~al}\mbox{.}(2013){Hacar}, {Tafalla}, {Kauffmann}, \&
  {Kov{\'a}cs}}]{haca13}
{Hacar} A., {Tafalla} M., {Kauffmann} J., {Kov{\'a}cs} A., 2013, \aap, 554, A55

\bibitem[{{Hills}(1980)}]{hill80}
{Hills} J.~G., 1980, \apj, 235, 986

\bibitem[{{H{\o}g} {et~al}\mbox{.}(2000){H{\o}g}, {Fabricius}, {Makarov},
  {Urban}, {Corbin}, {Wycoff}, {Bastian}, {Schwekendiek}, \& {Wicenec}}]{hog00}
{H{\o}g} E. {et~al.}, 2000, \aap, 355, L27

\bibitem[{{Hogg} {et~al}\mbox{.}(2010){Hogg}, {Bovy}, \& {Lang}}]{hogg10}
{Hogg} D.~W., {Bovy} J., {Lang} D., 2010, ArXiv e-prints

\bibitem[{{Holmberg} \& {Flynn}(2004)}]{holm04}
{Holmberg} J., {Flynn} C., 2004, \mnras, 352, 440

\bibitem[{{Hoogerwerf} {et~al}\mbox{.}(2000){Hoogerwerf}, {de Bruijne}, \& {de
  Zeeuw}}]{hoog00}
{Hoogerwerf} R., {de Bruijne} J.~H.~J., {de Zeeuw} P.~T., 2000, \apjl, 544,
  L133

\bibitem[{{Janson} {et~al}\mbox{.}(2013){Janson}, {Lafreni{\`e}re},
  {Jayawardhana}, {Bonavita}, {Girard}, {Brandeker}, \& {Gizis}}]{jans13}
{Janson} M., {Lafreni{\`e}re} D., {Jayawardhana} R., {Bonavita} M., {Girard}
  J.~H., {Brandeker} A., {Gizis} J.~E., 2013, \apj, 773, 170

\bibitem[{{Jeffries} {et~al}\mbox{.}(2014){Jeffries}, {Jackson}, {Cottaar},
  {Koposov}, {Lanzafame}, {Meyer}, {Prisinzano}, {Randich}, {Sacco},
  {Brugaletta}, {Caramazza}, {Damiani}, {Franciosini}, {Frasca}, {Gilmore},
  {Feltzing}, {Micela}, {Alfaro}, {Bensby}, {Pancino}, {Recio-Blanco}, {de
  Laverny}, {Lewis}, {Magrini}, {Morbidelli}, {Costado}, {Jofr{\'e}},
  {Klutsch}, {Lind}, \& {Maiorca}}]{jeff14}
{Jeffries} R.~D. {et~al.}, 2014, \aap, 563, A94

\bibitem[{{Jones}(1971)}]{jone71}
{Jones} D.~H.~P., 1971, \mnras, 152, 231

\bibitem[{{Karim} \& {Mamajek}(2017)}]{kari17}
{Karim} M.~T., {Mamajek} E.~E., 2017, \mnras, 465, 472

\bibitem[{{Kharchenko} {et~al}\mbox{.}(2007){Kharchenko}, {Scholz}, {Piskunov},
  {R{\"o}ser}, \& {Schilbach}}]{khar07}
{Kharchenko} N.~V., {Scholz} R.-D., {Piskunov} A.~E., {R{\"o}ser} S.,
  {Schilbach} E., 2007, Astronomische Nachrichten, 328, 889

\bibitem[{{Kouwenhoven} {et~al}\mbox{.}(2007){Kouwenhoven}, {Brown}, {Portegies
  Zwart}, \& {Kaper}}]{kouw07}
{Kouwenhoven} M.~B.~N., {Brown} A.~G.~A., {Portegies Zwart} S.~F., {Kaper} L.,
  2007, \aap, 474, 77

\bibitem[{{Kraus} {et~al}\mbox{.}(2015){Kraus}, {Cody}, {Covey}, {Rizzuto},
  {Mann}, \& {Ireland}}]{krau15}
{Kraus} A.~L., {Cody} A.~M., {Covey} K.~R., {Rizzuto} A.~C., {Mann} A.~W.,
  {Ireland} M.~J., 2015, \apj, 807, 3

\bibitem[{{Kroupa} {et~al}\mbox{.}(2001){Kroupa}, {Aarseth}, \&
  {Hurley}}]{krou01}
{Kroupa} P., {Aarseth} S., {Hurley} J., 2001, \mnras, 321, 699

\bibitem[{{Kroupa} {et~al}\mbox{.}(1999){Kroupa}, {Petr}, \&
  {McCaughrean}}]{krou99}
{Kroupa} P., {Petr} M.~G., {McCaughrean} M.~J., 1999, New Astronomy, 4, 495

\bibitem[{{Lada} \& {Lada}(2003)}]{lada03}
{Lada} C.~J., {Lada} E.~A., 2003, \araa, 41, 57

\bibitem[{{Lada} {et~al}\mbox{.}(1984){Lada}, {Margulis}, \&
  {Dearborn}}]{lada84}
{Lada} C.~J., {Margulis} M., {Dearborn} D., 1984, \apj, 285, 141

\bibitem[{{Larson}(1981)}]{lars81}
{Larson} R.~B., 1981, \mnras, 194, 809

\bibitem[{{Lepine} \& {Duvert}(1994)}]{lepi94}
{Lepine} J.~R.~D., {Duvert} G., 1994, \aap, 286, 60

\bibitem[{{Lindblad} {et~al}\mbox{.}(1973){Lindblad}, {Grape}, {Sandqvist}, \&
  {Schober}}]{lind73}
{Lindblad} P.~O., {Grape} K., {Sandqvist} A., {Schober} J., 1973, \aap, 24, 309

\bibitem[{{Lindegren} {et~al}\mbox{.}(2016){Lindegren}, {Lammers}, {Bastian},
  {Hern{\'a}ndez}, {Klioner}, {Hobbs}, {Bombrun}, {Michalik}, {Ramos-Lerate},
  {Butkevich}, {Comoretto}, {Joliet}, {Holl}, {Hutton}, {Parsons},
  {Steidelm{\"u}ller}, {Abbas}, {Altmann}, {Andrei}, {Anton}, {Bach},
  {Barache}, {Becciani}, {Berthier}, {Bianchi}, {Biermann}, {Bouquillon},
  {Bourda}, {Br{\"u}semeister}, {Bucciarelli}, {Busonero}, {Carlucci},
  {Casta{\~n}eda}, {Charlot}, {Clotet}, {Crosta}, {Davidson}, {de Felice},
  {Drimmel}, {Fabricius}, {Fienga}, {Figueras}, {Fraile}, {Gai}, {Garralda},
  {Geyer}, {Gonz{\'a}lez-Vidal}, {Guerra}, {Hambly}, {Hauser}, {Jordan},
  {Lattanzi}, {Lenhardt}, {Liao}, {L{\"o}ffler}, {McMillan}, {Mignard}, {Mora},
  {Morbidelli}, {Portell}, {Riva}, {Sarasso}, {Serraller}, {Siddiqui}, {Smart},
  {Spagna}, {Stampa}, {Steele}, {Taris}, {Torra}, {van Reeven}, {Vecchiato},
  {Zschocke}, {de Bruijne}, {Gracia}, {Raison}, {Lister}, {Marchant},
  {Messineo}, {Soffel}, {Osorio}, {de Torres}, \& {O'Mullane}}]{lind16}
{Lindegren} L. {et~al.}, 2016, \aap, 595, A4

\bibitem[{{Lodieu}(2013)}]{lodi13}
{Lodieu} N., 2013, \mnras, 431, 3222

\bibitem[{{Madsen} {et~al}\mbox{.}(2002){Madsen}, {Dravins}, \&
  {Lindegren}}]{mads02}
{Madsen} S., {Dravins} D., {Lindegren} L., 2002, \aap, 381, 446

\bibitem[{{Makarov}(2007)}]{maka07}
{Makarov} V.~V., 2007, \apjs, 169, 105

\bibitem[{{Mamajek} \& {Bell}(2014)}]{mama14}
{Mamajek} E.~E., {Bell} C.~P.~M., 2014, \mnras, 445, 2169

\bibitem[{{Mamajek} {et~al}\mbox{.}(2002){Mamajek}, {Meyer}, \&
  {Liebert}}]{mama02}
{Mamajek} E.~E., {Meyer} M.~R., {Liebert} J., 2002, \aj, 124, 1670

\bibitem[{{Marks} \& {Kroupa}(2012)}]{mark12}
{Marks} M., {Kroupa} P., 2012, \aap, 543, A8

\bibitem[{{Maschberger}(2013)}]{masc13}
{Maschberger} T., 2013, \mnras, 429, 1725

\bibitem[{{Michalik} {et~al}\mbox{.}(2015){Michalik}, {Lindegren}, \&
  {Hobbs}}]{mich15}
{Michalik} D., {Lindegren} L., {Hobbs} D., 2015, \aap, 574, A115

\bibitem[{{Moran}(1950)}]{mora50}
{Moran} P.~A.~P., 1950, Biometrika, 37, 17

\bibitem[{{Murphy} {et~al}\mbox{.}(2015){Murphy}, {Lawson}, \&
  {Bento}}]{murp15}
{Murphy} S.~J., {Lawson} W.~A., {Bento} J., 2015, \mnras, 453, 2220

\bibitem[{{O'dell} \& {Wen}(1994)}]{odel94}
{O'dell} C.~R., {Wen} Z., 1994, \apj, 436, 194

\bibitem[{{Odenkirchen} {et~al}\mbox{.}(2002){Odenkirchen}, {Grebel}, {Dehnen},
  {Rix}, \& {Cudworth}}]{oden02}
{Odenkirchen} M., {Grebel} E.~K., {Dehnen} W., {Rix} H.-W., {Cudworth} K.~M.,
  2002, \aj, 124, 1497

\bibitem[{{Olczak} {et~al}\mbox{.}(2008){Olczak}, {Pfalzner}, \&
  {Eckart}}]{olcz08}
{Olczak} C., {Pfalzner} S., {Eckart} A., 2008, \aap, 488, 191

\bibitem[{{Parker} \& {Goodwin}(2012)}]{park12d}
{Parker} R.~J., {Goodwin} S.~P., 2012, \mnras, 424, 272

\bibitem[{{Parker} \& {Quanz}(2012)}]{park12c}
{Parker} R.~J., {Quanz} S.~P., 2012, \mnras, 419, 2448

\bibitem[{{Parker} {et~al}\mbox{.}(2014){Parker}, {Wright}, {Goodwin}, \&
  {Meyer}}]{park14}
{Parker} R.~J., {Wright} N.~J., {Goodwin} S.~P., {Meyer} M.~R., 2014, \mnras,
  438, 620

\bibitem[{{Pecaut} \& {Mamajek}(2016)}]{peca16}
{Pecaut} M.~J., {Mamajek} E.~E., 2016, \mnras, 461, 794

\bibitem[{{Pecaut} {et~al}\mbox{.}(2012){Pecaut}, {Mamajek}, \&
  {Bubar}}]{peca12}
{Pecaut} M.~J., {Mamajek} E.~E., {Bubar} E.~J., 2012, \apj, 746, 154

\bibitem[{{Pfalzner}(2009)}]{pfal09}
{Pfalzner} S., 2009, \aap, 498, L37

\bibitem[{{Pfalzner} \& {Kaczmarek}(2013)}]{pfal13}
{Pfalzner} S., {Kaczmarek} T., 2013, \aap, 559, A38

\bibitem[{{Portegies Zwart} {et~al}\mbox{.}(2010){Portegies Zwart}, {McMillan},
  \& {Gieles}}]{port10}
{Portegies Zwart} S.~F., {McMillan} S.~L.~W., {Gieles} M., 2010, \araa, 48, 431

\bibitem[{{Preibisch} \& {Mamajek}(2008)}]{prei08}
{Preibisch} T., {Mamajek} E., 2008, {The Nearest OB Association:
  Scorpius-Centaurus (Sco OB2)}, {Handbook of Star Forming Regions}, p. 235

\bibitem[{{Preibisch} \& {Zinnecker}(1999)}]{prei99}
{Preibisch} T., {Zinnecker} H., 1999, \aj, 117, 2381

\bibitem[{{Rizzuto} {et~al}\mbox{.}(2011){Rizzuto}, {Ireland}, \&
  {Robertson}}]{rizz11}
{Rizzuto} A.~C., {Ireland} M.~J., {Robertson} J.~G., 2011, \mnras, 416, 3108

\bibitem[{{Rosotti} {et~al}\mbox{.}(2014){Rosotti}, {Dale}, {de Juan Ovelar},
  {Hubber}, {Kruijssen}, {Ercolano}, \& {Walch}}]{roso14}
{Rosotti} G.~P., {Dale} J.~E., {de Juan Ovelar} M., {Hubber} D.~A., {Kruijssen}
  J.~M.~D., {Ercolano} B., {Walch} S., 2014, \mnras, 441, 2094

\bibitem[{{Sartori} {et~al}\mbox{.}(2003){Sartori}, {L{\'e}pine}, \&
  {Dias}}]{sart03}
{Sartori} M.~J., {L{\'e}pine} J.~R.~D., {Dias} W.~S., 2003, \aap, 404, 913

\bibitem[{{Scally} \& {Clarke}(2001)}]{scal01}
{Scally} A., {Clarke} C., 2001, \mnras, 325, 449

\bibitem[{{Scally} \& {Clarke}(2002)}]{scal02}
{Scally} A., {Clarke} C., 2002, \mnras, 334, 156

\bibitem[{{Sch{\"o}nrich} {et~al}\mbox{.}(2010){Sch{\"o}nrich}, {Binney}, \&
  {Dehnen}}]{scho10}
{Sch{\"o}nrich} R., {Binney} J., {Dehnen} W., 2010, \mnras, 403, 1829

\bibitem[{{Taylor}(2005)}]{tayl05}
{Taylor} M.~B., 2005, in Astronomical Society of the Pacific Conference Series,
  Vol. 347, Astronomical Data Analysis Software and Systems XIV, {Shopbell} P.,
  {Britton} M., {Ebert} R., eds., p.~29

\bibitem[{{Tobin} {et~al}\mbox{.}(2015){Tobin}, {Hartmann}, {F{\H u}r{\'e}sz},
  {Hsu}, \& {Mateo}}]{tobi15}
{Tobin} J.~J., {Hartmann} L., {F{\H u}r{\'e}sz} G., {Hsu} W.-H., {Mateo} M.,
  2015, \aj, 149, 119

\bibitem[{{Tutukov}(1978)}]{tutu78}
{Tutukov} A.~V., 1978, \aap, 70, 57

\bibitem[{{van Leeuwen}(2007)}]{vanl07}
{van Leeuwen} F., 2007, \aap, 474, 653

\bibitem[{{Wright} {et~al}\mbox{.}(2016){Wright}, {Bouy}, {Drew}, {Sarro},
  {Bertin}, {Cuillandre}, \& {Barrado}}]{wrig16}
{Wright} N.~J., {Bouy} H., {Drew} J.~E., {Sarro} L.~M., {Bertin} E.,
  {Cuillandre} J.-C., {Barrado} D., 2016, \mnras, 460, 2593

\bibitem[{{Wright} {et~al}\mbox{.}(2012){Wright}, {Drake}, {Drew}, {Guarcello},
  {Gutermuth}, {Hora}, \& {Kraemer}}]{wrig12a}
{Wright} N.~J., {Drake} J.~J., {Drew} J.~E., {Guarcello} M.~G., {Gutermuth}
  R.~A., {Hora} J.~L., {Kraemer} K.~E., 2012, \apjl, 746, L21

\bibitem[{{Wright} {et~al}\mbox{.}(2014){Wright}, {Parker}, {Goodwin}, \&
  {Drake}}]{wrig14b}
{Wright} N.~J., {Parker} R.~J., {Goodwin} S.~P., {Drake} J.~J., 2014, \mnras,
  438, 639

\end{thebibliography}
\bsp

\end{document}